\documentclass[lettersize,journal]{IEEEtran}
\usepackage{cite}
\usepackage{amsmath,amssymb,amsfonts}
\usepackage{amsthm}
\usepackage{algorithmic}
\usepackage[ruled,vlined,linesnumbered]{algorithm2e}
\usepackage{graphicx}
\usepackage[caption=false,font=footnotesize,labelformat=simple]{subfig}

\usepackage{textcomp}
\usepackage{xcolor}
\usepackage{acronym}
\usepackage{bm}
\usepackage{booktabs}
\newtheorem{theorem}{Theorem}
\newtheorem{lemma}{Lemma}
\newtheorem{corollary}{Corollary}
\newtheorem{proposition}{Proposition}
\newtheorem{assumption}{Assumption}
\theoremstyle{remark}
\newtheorem{remark}{Remark}

\def\BibTeX{{\rm B\kern-.05em{\sc i\kern-.025em b}\kern-.08em
    T\kern-.1667em\lower.7ex\hbox{E}\kern-.125emX}}

\newcommand{\eest}{\epsilon_{\mathrm{est}}}
\newcommand{\eacc}{\epsilon_{\mathrm{inac}}}
\newcommand{\etot}{\epsilon_{\mathrm{tot}}}
\newcommand{\Etot}{E_{\mathrm{tot}}}
\newcommand{\Eagr}{E_{\mathrm{disag}}}
\newcommand{\Eacc}{E_{\mathrm{inac}}}
\newcommand{\Eest}{E_{\mathrm{est}}}
\newcommand{\phat}{\hat{\bm p}}
\newcommand{\ptil}{\tilde{\bm p}}
\newcommand{\pstar}{\bm p^\star}
\newcommand{\pbar}{\bar{\bm p}}
\newcommand{\pfix}{\bm p^\circ}
\newcommand{\pbeta}{\bm p_{\beta}}
\newcommand{\one}{\bm 1}
\newcommand{\xbar}{\bar x}
\newcommand{\betab}{\bm\beta}
\newcommand{\E}{\mathbb{E}}
\newcommand{\bmax}{\beta_{\max}}
\newcommand{\bmin}{\beta_{\min}}

\acrodef{mas}[MAS]{multi-agent system}
\acrodef{blue}[BLUE]{best linear unbiased estimate}
\acrodef{rms}[RMS]{root-mean-square}
\acrodef{jenc}[JENC]{joint estimation and network configuration}

\begin{document}

\setlength{\textfloatsep}{7pt plus 1pt minus 2pt}
\setlength{\floatsep}{5pt plus 1pt minus 2pt}
\setlength{\intextsep}{6pt plus 1pt minus 2pt}

\makeatletter
\def\ps@IEEEtitlepagestyle{%
    \def\@oddhead{\footnotesize\hfill
    This work has been submitted to IEEE for possible publication.
    Copyright may be transferred without notice, after which this
    version may no longer be accessible.\hfill}
    \def\@evenhead{\@oddhead}
    \def\@oddfoot{\hfill\thepage\hfill}
    \def\@evenfoot{\hfill\thepage\hfill}
}
\makeatother

\title{Joint Estimation and Network Configuration in Multi-Agent Systems}

\author{Mohammadreza Barzegaran,~\IEEEmembership{Member,~IEEE,} and Hamid Jafarkhani,~\IEEEmembership{Fellow,~IEEE}
\thanks{Mohammadreza Barzegaran and Hamid Jafarkhani are with the University of California, Irvine, CA 92697, USA\@. (e-mails: \{barzegm1, hamidj\}@uci.edu). This work was supported in part by the ARO grant W911NF2410046.}
}

\maketitle

\begin{abstract}
Distributed consensus-based sensing enables a multi-agent system to collaboratively estimate an unknown state by combining local measurements with information received from neighboring agents.
In practice, however, the communication network, transmission delays, and finite-rate communication directly affect the disagreement among agents.
On the other hand, agreement among agents does not ensure accuracy, particularly when sensors have different measurement uncertainties.
We develop a sampled consensus-plus-innovations framework under realistic wireless network constraints that jointly accounts for inter-agent disagreement and estimation inaccuracy.
We derive a computable upper bound on the worst-case steady-state estimation error that captures consensus disagreement due to the communication graph, quantization error, and transmission delay, together with estimation inaccuracy, which accounts for measurement uncertainty against the true state. 
We then use this bound as a network design objective and propose a semi-distributed auction-based algorithm that jointly modifies the communication graph and link-rate allocation.
The proposed method preserves connectivity, converges in finite time to a locally auction-stable network configuration, and reduces the bound on the estimation error at every accepted change.
The proposed method has been evaluated on various test cases, and the results show that it reduces the estimation error while maintaining fewer links and allocating moderate link rates in a short time.
\end{abstract}


\section{Introduction}
\Acp{mas} provide a scalable architecture for sensing, monitoring, and decision making without relying on a single high-quality sensor or a centralized fusion node.
This architecture is particularly useful for distributed sensing and coverage tasks such as cooperative target estimation, surveillance, environmental monitoring, and coverage control, where each agent only has a partial and noisy observation of the environment~\cite{olfati2007dkf,barzegaran2024framework,kar2012nonlinear,diaz2025multi,sayed2014adaptive,cortes2017coordinated,barzegaran2026encoded,barzegaran2025dynamic}.
Such cooperation is enabled by local communication, through which agents exchange information~\cite{barzegaran2026encoded,barzegaran2025dynamic,barzegaran2024framework}, and by consensus-based algorithms, through which they combine their information~\cite{olfati2007consensus,olfati2004time}.

In a consensus-based estimator, each agent repeatedly corrects its local estimate using information received from neighboring agents while incorporating its own measurement through an innovation term.
This is the \emph{consensus-plus-innovations} structure introduced by Kar and Moura~\cite{kar2009imperfect,kar2012nonlinear}.
Under ideal communication and contractive consensus dynamics, local estimates asymptotically approach a common value or converge to a bounded disagreement neighborhood around the true state. 
Classical consensus theory relates the speed and quality of this agreement to the spectral properties of the graph Laplacian, particularly the algebraic connectivity of the network~\cite{olfati2007consensus,fiedler1973algebraic, jadbabaie2003coordination,fax2004information}. 
However, under realistic communication scenarios, finite link rates introduce quantization error~\cite{kashyap2007quantized,nedic2009quantization,10106411}, transmission introduces delay~\cite{olfati2004time,liu2014distributed,ortega2024decentralized}, and the limited communication range restricts feasible links~\cite{barzegaran2024framework,barzegaran2025dynamic}.
These communication constraints affect both the dynamics of the estimator and the resources required to exchange information.
Therefore, estimation performance and network configuration must be treated as a jointly constrained design problem~\cite{cortes2017coordinated,chen2019bit}.

Quantized and delayed consensus have been studied extensively, both separately and jointly~\cite{zhang2024bipartite,niu2023robust,chen2019bit,li2010distributed,liu2014distributed,olfati2004time,ghosh2006growing,cortes2017coordinated}.
These works analyze how closely the agents agree, but do not determine how close the agreed value is to the true state.
For example, a homogeneous innovation gain applied to measurements with different uncertainties may converge to a common consensus value with a large inaccuracy.
In contrast, this work considers a local estimation error at each agent with two components: consensus disagreement and estimation inaccuracy against the true state.
Taking into account both components, we derive a computable upper bound on the worst-case steady-state estimation error.
In this bound, the disagreement part drives the network design, whereas the inaccuracy part selects the innovation gains and certifies the accuracy against the true state.
Since the bound is a worst-case guarantee, we separately evaluate how tightly it certifies the error and how well it ranks network configurations.
In addition, we turn this \emph{estimation error bound} into a design cost function for the joint selection of the communication topology and the finite-rate allocation.
This connects estimator analysis with the topology-design literature, which maximizes algebraic connectivity or grows well-connected graphs~\cite{kim2006lambda2,ghosh2006growing,cortes2017coordinated,kar2008randomlinks}.
Unlike that literature, our bound analysis also accounts for finite link rates and delay, and the topology is configured by
a semi-distributed configuration procedure rather than a centralized graph search.
Existing estimators analyze agreement on a fixed graph. Spectral topology design approaches ignore estimation accuracy and finite link rates.
In contrast, our method bounds both disagreement and inaccuracy under quantization and delay, and uses this bound to design the topology, link rates, and consensus gain.
Table~\ref{tab:related_work} compares our proposed method with related approaches.

The main contributions of this paper are: 
\begin{itemize}
\item A consensus-plus-innovations estimator with standard inverse-variance innovation gains~\cite{kay1993estimation} is formulated (Lemma~\ref{lem:fixed}).
\item A computable worst-case steady-state bound on the estimation error that captures the network graph, quantization, delay, and measurement uncertainty is derived. Unlike staleness-based bounds, the communication term of this bound does not grow with delays (Theorem~\ref{thm:total} and Lemma~\ref{lem:wc}).
\item The effects of the innovation gain profile and innovation scale on the fusion, communication, and disagreement terms of this bound are established (Propositions~\ref{prop:structure}--\ref{prop:scale} and Corollary~\ref{cor:spectral}).
\item A semi-distributed auction-based algorithm is proposed to configure the network graph, link rates, and consensus gain. 
\item A proof of finite termination without the sweep cap and that, upon termination, no score-qualified single-link action passes the verification of the coordinator is provided (Proposition~\ref{prop:auction}).
\end{itemize}
\begin{table}[t]
    \centering
    \footnotesize
    \caption{Related Work: ``Fixed'' Indicates That the Graph Is Analyzed but Not Designed.}
    \label{tab:related_work}
    \resizebox{\columnwidth}{!}{%
    \begin{tabular}{lccccc}
        \toprule
        Approach & Disagreement & Inaccuracy & Quant./delay & Topology & Rates \\
        \midrule
        Consensus+innovations~\cite{kar2009imperfect,kar2010quantized,kar2012nonlinear} & Yes & Asymptotic & Quant. & Fixed & No \\
        Consensus fusion~\cite{xiao2005fusion,xiao2007lms} & Yes & BLUE & Noise & Fixed & No \\
        Quantized/delayed consensus~\cite{nedic2009quantization,aysal2008dithered,olfati2004time,chen2019bit} & Yes & No & Both & Fixed & Conditions \\
        Spectral topology design~\cite{ghosh2006growing,kar2008randomlinks} & Conv.\ rate & No & Link failures & Designed & Link use \\
        This work & Bounded & Bounded & Both & Designed & Designed \\
        \bottomrule
    \end{tabular}}
\end{table}

The remainder of the paper is organized as follows.
The system model and the problem statement are explained in Section~\ref{sec:problemdef}.
The estimator framework and the estimation error bound are presented in Section~\ref{sec:framework}, and the network configuration algorithm is described in Section~\ref{sec:netdesign}.
The evaluation results are presented in Section~\ref{sec:evaluation}, and the concluding remarks are provided in Section~\ref{sec:conclusion}.
Proofs are given in the appendix.

\emph{Notation:} Lowercase letters denote scalars, boldface lowercase letters denote vectors, uppercase letters denote matrices, and script letters denote sets.
Hats denote estimates. The symbol $\otimes$ denotes the Kronecker product, $\one$ is the all-ones vector, $I$ is an identity matrix of the required dimension, and $\Pi=I-\tfrac1N\one\one^\top$ subtracts the network average from each entry.
For a vector $\bm y$, $\mathrm{diag}(\bm y)$ denotes the diagonal matrix with diagonal $\bm y$.
Vector and matrix inequalities hold entrywise.
The norm $\|\cdot\|$ denotes the Euclidean norm for vectors and the induced spectral norm for matrices.
The expectation $\E$ is taken over measurement uncertainty.
The uppercase $E$ denotes a population \ac{rms} performance measure and the lowercase $\epsilon$ denotes its analytical bound.

\section{Problem Statement}\label{sec:problemdef}
Consider an \ac{mas} consisting of $N\ge3$ heterogeneous static agents indexed by $\mathcal N=\{1,\ldots,N\}$.
The agents cooperatively estimate an unknown static state $\pstar\in\mathbb R^r$, where $r\ge1$ is the dimension of the state.
Each agent~$n$ is equipped with a sensor that returns a single snapshot of an uncertain measurement,
\begin{align}
    \ptil_n=\pstar+\bm v_n,
    \quad \E\bm v_n=\bm 0,\ \E\bm v_n\bm v_n^\top=\sigma_n^2I_r,
    \label{eq:obs_model}
\end{align}
where the uncertainty vectors $\bm v_n$ are mutually independent with known variances $\sigma_n^2>0$.
An example of such sensors is a range-bearing sensor with range-dependent sensing noise~\cite{blanco2008rangebearing}.

Agents communicate over an undirected graph $\mathcal{G}(\mathcal{N},\mathcal{E})$, where $(n,m)\in\mathcal{E}$ if there is a link between Agents~$n$ and~$m$.
The neighbor set of Agent~$n$ is $\mathcal{N}_n=\{m\in\mathcal{N}:(n,m)\in\mathcal{E}\}$.
Let $A=[a_{n,m}]$ be the adjacency matrix, where $a_{n,m}=1$ if $(n,m)\in\mathcal{E}$ and $a_{n,m}=0$ otherwise. Let $D=\operatorname{diag}(d_1,\ldots,d_N)$ be the degree matrix, with $d_n=\sum_{m\in\mathcal{N}}a_{n,m}$, and let $L=D-A$ be the graph Laplacian.

Communication is subject to delay and quantization because each link can transmit only at a finite bit-rate, dictated by its capacity and other practical limitations.
Let $k$ and $T_s$ denote the discrete time index and the time-slot duration, respectively.
Each link $(n,m)\in\mathcal{E}$ has a bounded integer delay $0\le\tau_{n,m}\le\tau_{\max}$.
When Agent~$n$ transmits a symbol $\bm{z}[k]$ at time $k$ to Agent~$m$, the symbol is received at time $k+\tau_{n,m}$.
Therefore, Agent~$m$ receives the delayed symbol $\bm{z}[k-\tau_{n,m}]$ rather than the current symbol $\bm{z}[k]$.

Additionally, the link $(n,m)$ has a finite bit-rate $R_{n,m}$, which limits the number of bits that can be transmitted.
Taking into account the update period $T_s$, the per-coordinate resolution is $b_{n,m}=b_{m,n}=\lfloor R_{n,m}T_s/r\rfloor$.
The allocation satisfies the per-agent budgets $\sum_mb_{n,m}\le\bar B_n$ and the network-wide budget $\sum_{(n,m)\in\mathcal E}b_{n,m}\le B_{\mathrm{tot}}$, with each link counted once.
Messages are encoded by a uniform quantizer $Q_b$ whose spacing corresponds to $b$ bits per coordinate.
The corresponding quantizer satisfies
\begin{equation}
    Q_{b_{n,m}}(\bm{z})=\bm{z}+\bm{\epsilon}^{q}_{n,m},
    \qquad
    \|\bm{\epsilon}^{q}_{n,m}\|\le c2^{-b_{n,m}},
    \label{eq:quantization}
\end{equation}
for a quantizer-dependent constant $c>0$~\cite{720541}.
Here, $\bm{\epsilon}^{q}_{n,m}$ is the quantization error on link~$(n,m)$.

Combining delay and quantization, let $\bm{q}_{n,m}$ denote the delayed and quantized estimate of Agent~$m$ available to Agent~$n$ over link~$(n,m)$.
With \eqref{eq:quantization}, this can be defined as
\begin{align}
    \label{eq:delayedquantized}
    \nonumber \bm{q}_{n,m}[k]&=Q_{b_{n,m}}(\phat_m[k-\tau_{n,m}])\\
    &=\phat_m[k-\tau_{n,m}] + \bm{\epsilon}^{q}_{n,m},
\end{align}
where $\phat_m[k-\tau_{n,m}]\in\mathbb R^r$ denotes the local estimate at Agent~$m$ at time step~$k-\tau_{n,m}$. This is the message that Agent~$m$ transmits to Agent~$n$ and Agent~$n$ receives it at time step~$k$.
Note that~\eqref{eq:quantization} holds only for inputs within the range of the quantizer.
For a uniform quantizer with $b$ bits per coordinate and a range $[-\kappa,\kappa]$ around a known center in each coordinate, the spacing between adjacent levels is $2\kappa\,2^{-b}$ and $c=\sqrt r\,\kappa$.
Therefore, the dimension $r$ affects~\eqref{eq:quantization} through both $c$ and $b_{n,m}$.

Assume that the initial network configuration is valid and is represented by a connected graph $\mathcal{G}_0(\mathcal{N},\mathcal{E}_0)$ with an initial bit-rate allocation $\mathcal{B}_0$.
The objective is to co-design a sampled distributed consensus estimator that computes the local estimate~$\phat_n$ at Agent~$n$, and a network configuration $(\mathcal{G},\mathcal{B})$ consisting of the communication graph and bit-rate allocation.
Let the worst-case steady-state estimation error be
\begin{align}
    \label{eq:Etot}
    \Etot&=\max_{n \in \mathcal{N}}\limsup_{k\to\infty}\big(\E\|\phat_n[k]-\pstar\|^2\big)^{1/2}.
\end{align}
The design objective is to reduce $\Etot$ despite delay, quantization, and measurement uncertainty.

\section{Consensus Framework}\label{sec:framework}

The proposed framework consists of a distributed consensus estimator in which each agent maintains a local estimate of the unknown state.
Specifically, Agent~$n$ maintains a local estimate $\phat_n[k]$ at a discrete time step $k$ and updates it using the consensus-plus-innovations structure~\cite{kar2010quantized,kar2012nonlinear},
\begin{align}
    \nonumber \phat_n[k+1]=\phat_n[k]&-\alpha\!\sum_{m\in\mathcal N_n}\!a_{n,m}\big(\phat_n[k]-\bm q_{n,m}[k]\big)\\
    &+\beta_n\big(\ptil_n-\phat_n[k]\big),
    \label{eq:estimator}
\end{align}
where $\alpha>0$ is the consensus gain, $\beta_n$ is the innovation gain of Agent~$n$, and $\bm q_{n,m}[k]$ is the delayed and quantized estimate in~\eqref{eq:delayedquantized}.
The consensus term reduces the disagreement with neighboring estimates, whereas the innovation term pulls the estimate toward the local measurement.
Since $\bm q_{n,m}[k]$ reads $\phat_m[k-\tau_{n,m}]$, the recursion requires a history of negative indices during the first $\tau_{\max}=\max_{(n,m)}\tau_{n,m}$ iterations.
For these negative indices, the history is set as $\phat_n[k]=\ptil_n$ for $-\tau_{\max}\le k\le0$.

Since sensors have different uncertainties, the innovation gains $\beta_n$ should not be equal.
They are selected to reflect the relative certainty of each measurement.
We select innovation gains according to the inverse-variance profile as
\begin{equation}
    \beta_n=\beta_0\,\frac{\sigma_n^{-2}}{\max_m\sigma_m^{-2}},\qquad \beta_0\in(0,1),
    \label{eq:gains}
\end{equation}
so that the most certain sensor receives the largest gain $\beta_0$. Let $\betab=[\beta_1,\ldots,\beta_N]^\top$ collect the innovation gains, and let $\bmin$ and $\bmax$ denote the smallest and largest gains, respectively. Note that under~\eqref{eq:gains}, $\bmax=\beta_0$.
This is motivated by the \ac{blue} for independent measurements with heterogeneous uncertainties~\cite{kay1993estimation,xiao2005fusion}.
For measurements $\ptil_1,\ldots,\ptil_N$, the common \ac{blue} and its \ac{rms} error are, respectively,
\begin{equation}
\begin{aligned}
    \bm p_{\mathrm{BLUE}}&=\frac{\sum_n\sigma_n^{-2}\ptil_n}{\sum_n\sigma_n^{-2}},\\
    \epsilon_{\mathrm{BLUE}}&=\Big(\E\|\bm p_{\mathrm{BLUE}}-\pstar\|^2\Big)^{1/2}=\sqrt{\frac{r}{\sum_n\sigma_n^{-2}}},
\end{aligned}
    \label{eq:blue}
\end{equation}
which achieves the minimum mean-square error among all unbiased linear fusion rules~\cite{kay1993estimation}.
In~\eqref{eq:estimator}, Agent~$n$ weights its own estimate by $1-\alpha d_n-\beta_n$, which becomes negative if the consensus gain is too large relative to the degree $d_n$ and the innovation gain $\beta_n$.
To keep this weight nonnegative, we restrict the gains as follows.

\begin{assumption}[Positivity]\label{ass:step}
The gains satisfy $\beta_n\in(0,1)$ and $\alpha d_n+\beta_n\le1$ for every $n \in \mathcal{N}$. Equivalently, $0<\alpha\le\bar\alpha(L,\betab):=\min_n(1-\beta_n)/d_n$.
\end{assumption}

Under Assumption~\ref{ass:step}, the local estimator in~\eqref{eq:estimator} is the convex combination
\begin{align}
    \nonumber &\phat_n[k+1]=\\
    &(1-\alpha d_n-\beta_n)\phat_n[k]+\alpha\sum_m a_{n,m}\bm q_{n,m}[k]+\beta_n\ptil_n.
    \label{eq:convex}
\end{align}
Since all weights in~\eqref{eq:convex} are nonnegative, the delayed estimates can be bounded through a window of previous errors.
Moreover, the iterates remain bounded under quantization and unlike in some quantized consensus schemes, they need not be projected onto the range of the quantizer.
For the analysis, let $B=\mathrm{diag}(\betab)$ and define
\begin{equation}
    W=I-B-\alpha L,\qquad G=(B+\alpha L)^{-1},\qquad H=GB.
    \label{eq:matrices}
\end{equation}
The matrix $W$ is symmetric, all its entries are nonnegative, and the sum of its $n$-th row is $1-\beta_n$. 
Additionally, $\tilde{\bm P}$ and $\bm V$ stack the measurements $\ptil_n$ and the uncertainty vectors $\bm v_n$, respectively, and $S=\mathrm{diag}(\sigma_1^2,\ldots,\sigma_N^2)$ collects the measurement variances.

\begin{lemma}[Ideal consensus value]\label{lem:fixed}
Under Assumption~\ref{ass:step}, $G\ge0$, $H\ge0$, $H\one=\one$, and $\betab^\top H=\betab^\top$. Without delay and quantization, the recursion~\eqref{eq:estimator} has a unique fixed point $\bm P^\circ=(H\otimes I_r)\tilde{\bm P}$ and converges to $\bm P^\circ$ under arbitrary bounded delays. Each $\pfix_n$ is a convex combination of the measurements, and
\begin{align}
    \small
    \sum_n\beta_n\pfix_n=\sum_n\beta_n\ptil_n
    \;\Longrightarrow\;
    \frac{\sum_n\beta_n\pfix_n}{\sum_n\beta_n}=\pbeta:=\frac{\sum_n\beta_n\ptil_n}{\sum_n\beta_n}.
    \label{eq:weighted_mean}
\end{align}
Under the innovation gains in~\eqref{eq:gains}, $\pbeta=\bm p_{\mathrm{BLUE}}$.
\end{lemma}

Lemma~\ref{lem:fixed} relates the distributed estimator to the centralized fusion.
Under the innovation gains in~\eqref{eq:gains}, the gain-weighted mean of the ideal consensus value is equal to that of \ac{blue}.
This does not imply that the agents reach the same value because the innovation term keeps each estimate closer to its own measurement.
Moreover, with heterogeneous gains, the unweighted network average of the ideal consensus value can differ from the gain-weighted fusion $\pbeta$.
Therefore, the bound must account for these two ideal errors and for the perturbation introduced by delayed and quantized communication.


In practice, each agent receives delayed and quantized estimates. Therefore, its local estimate deviates from the ideal consensus value.
To quantify this deviation, let $\bm x_n[k]=\phat_n[k]-\pfix_n$ denote the deviation of Agent~$n$ from the ideal consensus value at time step~$k$.

The deviation is driven by the quantization error on each link.
From~\eqref{eq:delayedquantized}, the quantization error at time step~$k$ is $\bm\epsilon^q_{n,m}[k]=\bm q_{n,m}[k]-\phat_m[k-\tau_{n,m}]$, and from~\eqref{eq:quantization}, it satisfies $\|\bm\epsilon^q_{n,m}[k]\|\le c\,2^{-b_{n,m}}$.
Since this bound holds for every input, no statistical assumption on the quantization error is required.
Additionally, Agent~$n$ combines the estimates of its neighbors with the consensus gain $\alpha$.
Therefore, with~\eqref{eq:convex}, the quantization 
adds the term $\alpha\sum_m a_{n,m}\bm\epsilon^q_{n,m}[k]$ to its update. The norm of this term
is bounded by the per-agent quantization load
\begin{equation}
    q_n=\alpha\!\sum_{m}a_{n,m}c\,2^{-b_{n,m}}.
    \label{eq:loads}
\end{equation}
Note that $q_n$ increases with the consensus gain and the number of neighbors and decreases exponentially with the link resolutions.

Unlike quantization, delay does not introduce a separate error term.
Instead, it determines which past estimates are combined in each update.
Under Assumption~\ref{ass:step}, each update in~\eqref{eq:convex} is a convex combination of the current estimate, the delayed estimates of neighbors, and the measurement.
Therefore, delayed estimates cannot move the deviation beyond the largest deviation in the recent past, except through the bounded quantization error~\cite{haddad2004delay}.
The following lemma uses this property to bound the deviation for arbitrary bounded delays.

\begin{lemma}[Communication perturbation]\label{lem:wc}
Let Assumption~\ref{ass:step} hold and define $\psi_0=\max_{n,\,-\tau_{\max}\le j\le0}\|\bm x_n[j]\|$. Then, for arbitrary delays $0\le\tau_{n,m}\le\tau_{\max}$, every realization, and every $k\ge1$,
\begin{equation}
    \max_n\|\bm x_n[k]\|\le\frac{\max_nq_n}{\bmin}+(1-\bmin)^{\lfloor k/(\tau_{\max}+1)\rfloor}\psi_0,
    \label{eq:transient}
\end{equation}
and, for every agent~$n$,
\begin{equation}
    \limsup_{k\to\infty}\|\bm x_n[k]\|\le\xbar_n:=[G\bm q]_n.
    \label{eq:xbar_wc}
\end{equation}
\end{lemma}

Lemma~\ref{lem:wc} provides two bounds.
The first bound~\eqref{eq:transient} describes the transient behavior.
The effect of the initial deviation $\psi_0$ decays by a factor of $1-\bmin$ every $\tau_{\max}+1$ time steps.
Therefore, longer delays slow down the convergence.
The second bound~\eqref{eq:xbar_wc} describes the steady state.
It depends on the graph, the gains, and the link resolutions through $G$ and $\bm q$, but not on the delays.
Therefore, delay affects how fast the estimator converges, but not how close it remains to the ideal consensus value.

Additionally, Lemma~\ref{lem:wc} determines the range of the quantizer.
Since each $\pfix_n$ is a convex combination of the measurements, every estimate lies within $\max_mq_m/\bmin+\psi_0$ of the convex hull of the measurements, where $\psi_0$ is at most the diameter of this hull.
Note that the bound~\eqref{eq:transient} at time step $k+1$ uses the quantization errors only up to time step $k$.
Therefore, by induction, if the quantizer range covers this set, the quantizer never overloads and~\eqref{eq:quantization} is in effect at every time step.
If the measurements are within a known bounded set, e.g., the sensing range, the range can be selected before deployment whenever $\max_nq_n<\kappa\,\bmin$, because $q_n$ also grows with $\kappa$.

\begin{remark}[Staleness-based bounds]
A common alternative treats delay as staleness, i.e., it bounds $\|\phat_m[k-\tau]-\phat_m[k]\|$ by $\tau$ times the largest change per update and adds this quantity as a disturbance. This bound grows linearly with the delay, although the lossless estimator converges to the same fixed point for every delay. Lemma~\ref{lem:wc} avoids this term by directly analyzing the delayed recursion.
\end{remark}

We now combine the results of Lemma~\ref{lem:fixed} and Lemma~\ref{lem:wc} into a bound on the estimation error~\eqref{eq:Etot}.
Let $\pbar[k]=\tfrac1N\sum_n\phat_n[k]$ denote the network average of the local estimates.
Since
\begin{align}
    \phat_n[k]-\pstar=(\phat_n[k]-\pbar[k])+(\pbar[k]-\pstar),
\end{align}
Minkowski's inequality and the subadditivity of the $\limsup$~\cite{stein2009real} result in
\begin{align}
    \Etot \le \Eagr+\Eacc,
    \label{eq:split}
\end{align}
where
\begin{align}
    \Eagr&=\max_n\limsup_{k\to\infty}\big(\E\|\phat_n[k]-\pbar[k]\|^2\big)^{1/2},\label{eq:Eagr}
\end{align}
is the disagreement, i.e., the largest deviation of an agent from the network average, and
\begin{align}
    \Eacc&=\limsup_{k\to\infty}\big(\E\|\pbar[k]-\pstar\|^2\big)^{1/2},\label{eq:Eacc}
\end{align}
is the inaccuracy, i.e., the error of the network average against the true state.
Therefore, the estimation error of each agent contains two components: disagreement and inaccuracy.
Evaluating $\Etot$ directly requires the distribution of the estimation error and the steady-state trajectory of the network.
Therefore, we bound $\Eagr$ and $\Eacc$ separately by computable bounds.


We first bound the disagreement~$\Eagr$.
Without delay and quantization, the estimates converge to the ideal consensus value in Lemma~\ref{lem:fixed}.
Since using  \eqref{eq:estimator}, each agent is pulled towards its own measurement, the ideal estimates disagree.
The \ac{rms} distance of Agent~$n$ from the network average of the ideal consensus value is
\begin{equation}
    \epsilon_{\mathrm{dis},n}=\sqrt{r\,[\Pi HSH^\top\Pi]_{nn}},
    \label{eq:edis}
\end{equation}
where $\Pi$ subtracts the network average from each entry.
Additionally, from Lemma~\ref{lem:wc}, the delayed and quantized communication moves the estimate of Agent~$n$ away from the ideal consensus value by at most $\xbar_n$ in steady state, and moves the network average by at most $\tfrac1N\sum_m\xbar_m$.
Therefore, the disagreement is bounded by
\begin{equation}
    \tfrac12\eest=\max_n\Big[\epsilon_{\mathrm{dis},n}+\big(1-\tfrac2N\big)\xbar_n+\tfrac1N\textstyle\sum_m\xbar_m\Big],
    \label{eq:eest}
\end{equation}
where the last two terms bound the communication deviation of Agent~$n$ relative to the network average.

Note that $\tfrac12$ in \eqref{eq:eest} ensures that $\eest$ bounds the disagreement between any two agents, which is at most twice the disagreement of an agent from the network average, as shown next.
Let
\begin{equation}
    \Eest=\max_{n,m}\limsup_{k\to\infty}\big(\E\|\phat_n[k]-\phat_m[k]\|^2\big)^{1/2}
    \label{eq:Eest}
\end{equation}
denote the largest disagreement between two agents.
Since $\phat_n[k]-\phat_m[k]=(\phat_n[k]-\pbar[k])-(\phat_m[k]-\pbar[k])$, Minkowski's inequality gives~\cite{stein2009real}
\begin{equation}
    \Eest\le2\Eagr.
    \label{eq:pairwise}
\end{equation}
Therefore, $\Eest\le\eest$.

Next, we bound the inaccuracy~$\Eacc$.
The error of the network average against the true state has three components.
First, from Lemma~\ref{lem:fixed}, the gain-weighted mean of the ideal consensus value is equal to the gain-weighted fusion $\pbeta$, which retains the measurement uncertainty.
Second, with heterogeneous gains, the unweighted network average of the ideal consensus value differs from $\pbeta$.
Third, the delayed and quantized communication moves the network average by at most $\tfrac1N\sum_m\xbar_m$.
The \ac{rms} error of the fusion and the \ac{rms} difference between the two network averages are
\begin{equation}
    \epsilon_{\mathrm{fus}}=\frac{\sqrt{r\sum_n\beta_n^2\sigma_n^2}}{\sum_n\beta_n},\qquad
    \epsilon_{\mathrm{gs}}=\sqrt{r\,\bm w^\top S\bm w},
    \label{eq:efus}
\end{equation}
respectively, where $\bm w=\tfrac1NH^\top\one-\betab/(\one^\top\betab)$.
Therefore, the inaccuracy term is bounded by
\begin{equation}
    \eacc=\epsilon_{\mathrm{fus}}+\epsilon_{\mathrm{gs}}+\tfrac1N\textstyle\sum_m\xbar_m.
    \label{eq:eacc}
\end{equation}

Finally, combining the disagreement and inaccuracy bounds with~\eqref{eq:split}, the total estimation error bound is
\begin{equation}
    \etot=\tfrac12\eest+\eacc.
    \label{eq:etot}
\end{equation}

\begin{theorem}[Total-error bound]\label{thm:total}
Under the model of Section~\ref{sec:problemdef} and Assumption~\ref{ass:step}, let $\xbar_n$ be given by~\eqref{eq:xbar_wc}. Then, for arbitrary delays $0\le\tau_{n,m}\le\tau_{\max}$,
\begin{align}
    \Eest\le\eest,\quad \Eagr\le\tfrac12\eest,\quad \Eacc\le\eacc,\quad \Etot\le\etot.
    \label{eq:certificate}
\end{align}
Every term in~\eqref{eq:edis}--\eqref{eq:etot} is an explicit function of $(L,\mathcal B,\alpha,\betab,S,c)$; neither the measurements, nor the trajectory, nor the delay values are required.
\end{theorem}



\subsection{Effect of the Innovation Gains}
\label{sec:innogains}
The innovation gains in~\eqref{eq:gains} have two parts: a profile and a scale.
The profile $\sigma_n^{-2}/\max_m\sigma_m^{-2}$ determines the relative weight of each measurement, and the scale $\beta_0$ determines how strongly each agent is attracted to its own measurement.
Both parts affect the bound~\eqref{eq:etot}.
We first study the effect of the profile on the inaccuracy bound~\eqref{eq:eacc}. Let $\bar\beta=\tfrac1N\sum_n\beta_n$ denote the mean innovation gain.

\begin{proposition}[Fusion, gain spread, and communication]\label{prop:structure}
Under the assumptions of Theorem~\ref{thm:total}, the following statements hold.
\begin{itemize}
    \item[(a)] $\epsilon_{\mathrm{fus}}\ge\epsilon_{\mathrm{BLUE}}$, with equality if and only if $\beta_n\propto\sigma_n^{-2}$, i.e., under~\eqref{eq:gains}.
    \item[(b)] The ideal network average satisfies $\bar{\bm p}^\circ-\pbeta=-\frac1{N\bar\beta}\sum_n(\beta_n-\bar\beta)(\pfix_n-\bar{\bm p}^\circ)$. Consequently,
    \begin{equation}
        \epsilon_{\mathrm{gs}}\le\frac{\|\betab-\bar\beta\one\|}{N\bar\beta}\Big(\textstyle\sum_n\epsilon_{\mathrm{dis},n}^2\Big)^{1/2},
        \label{eq:gain_spread}
    \end{equation}
    and $\epsilon_{\mathrm{gs}}=0$ for homogeneous gains.
    \item[(c)] $\sum_n\beta_n\xbar_n=\sum_nq_n$. With $\bar\epsilon_c=\tfrac1N\sum_nq_n$, the gain-weighted mean of $\xbar$ is equal to $\bar\epsilon_c/\bar\beta$, and $\tfrac1N\sum_n\xbar_n\ge\bar\epsilon_c/\bmax$, with the equality $\bar\epsilon_c/\beta$ for homogeneous gains $\beta_n\equiv\beta$.
\end{itemize}
\end{proposition}

Proposition~\ref{prop:structure} shows that the choice of profile is a trade-off.
From Statement~(a), the gains in~\eqref{eq:gains} minimize the fusion error, whereas homogeneous gains have a larger fusion error when the sensor variances differ.
From Statement~(b), homogeneous gains eliminate the gain-spread error, whereas heterogeneous gains shift the network average away from $\pbeta$.
However, this shift is bounded by the ideal disagreement in~\eqref{eq:gain_spread}.
Therefore, a network with a small disagreement also has a small gain-spread error.
From Statement~(c), the contribution of communication to inaccuracy is at least $\bar\epsilon_c/\bmax$.
Therefore, a weaker innovation leaves more communication error in the network average.

Next, with the gain profile fixed, we study the effect of the scale $\beta_0$.
A smaller scale allows the consensus term to combine the measurements more completely.
However, it also weakens the innovation term, which dampens communication errors.

\begin{proposition}[Innovation scale]\label{prop:scale}
Let the graph be connected, fix the profile $\betab=\beta_0\tilde{\bm w}$ with $\max_n\tilde w_n=1$, and fix a consensus gain $\alpha$ that is admissible for all $\beta_0\le\beta_0'$. As $\beta_0\to0$, $H\to\one\tilde{\bm w}^\top/(\one^\top\tilde{\bm w})$. Therefore, $\epsilon_{\mathrm{dis},n}\to0$ and $\epsilon_{\mathrm{gs}}\to0$, whereas $\epsilon_{\mathrm{fus}}$ does not depend on $\beta_0$. Consequently, the following statements hold.
\begin{itemize}
    \item[(a)] For lossless links ($q_n=0$), $\etot\to\epsilon_{\mathrm{fus}}$, which equals $\epsilon_{\mathrm{BLUE}}$ under~\eqref{eq:gains}.
    \item[(b)] For lossy links, $\eacc\ge\bar\epsilon_c/\beta_0$; therefore, $\etot\to\infty$, and a minimum is attained at a positive scale on the admissible range $(0,\beta_0']$.
\end{itemize}
\end{proposition}

Proposition~\ref{prop:scale} shows that the scale $\beta_0$ should not be too small.
Over lossless links, decreasing $\beta_0$ drives the estimates toward the gain-weighted fusion, which is the same as \ac{blue} under~\eqref{eq:gains}.
Over quantized links, decreasing $\beta_0$ increases the communication contribution without bound.
Additionally, a smaller scale slows down the convergence, because the transient bound~\eqref{eq:transient} decays with $\bmin$.
Therefore, the scale $\beta_0$ trades off the fusion of the measurements against the communication error and the convergence time.
This trade-off also motivates the innovation term over the ratio consensus, i.e., consensus on $\sigma_n^{-2}\ptil_n$ and $\sigma_n^{-2}$ followed by their ratio.
Over lossless links, the ratio consensus converges to the \ac{blue} with zero disagreement~\cite{xiao2005fusion}.
However, it is not anchored to the measurements, so quantization errors accumulate in the network sums.
In contrast, the innovation term keeps the communication perturbation bounded (Lemma~\ref{lem:wc}) at the cost of a nonzero ideal disagreement and extends directly to fresh measurements~\cite{kar2012nonlinear}. We have further investigated this motivation numerically in Section~\ref{sec:evaluation}.

\subsection{Spectral Interpretation}
In classical consensus, disagreement among agents is related to the algebraic connectivity $\lambda_2(L)$, i.e., the second-smallest eigenvalue of $L$~\cite{olfati2007consensus,fiedler1973algebraic}.
However, the ideal disagreement~\eqref{eq:edis} does not show this relation explicitly.
Therefore, we also bound it through the spectrum of $W$.
To this end, we define
\begin{equation}
    \rho=\|U^\top WU\|,\quad g=\|\betab-\bar\beta\one\|,\quad \Delta=(1-\rho)\bar\beta-\tfrac{g^2}{N},
    \label{eq:rho}
\end{equation}
where $U\in\mathbb R^{N\times(N-1)}$ has orthonormal columns that span $\one^\perp$.
Here, $\rho$ is the contraction factor of $W$ on the disagreement subspace $\one^\perp$, and $g$ measures the spread of the innovation gains.

\begin{corollary}[Spectral bound]\label{cor:spectral}
If $\Delta>0$, then
\begin{equation}
    \Big(\textstyle\sum_n\epsilon_{\mathrm{dis},n}^2\Big)^{1/2}\le\frac{\bar\beta\,\bar d_s}{\Delta},\quad
    \bar d_s^2=\E\textstyle\sum_n\beta_n^2\|\ptil_n-\pbeta\|^2.
    \label{eq:spectral}
\end{equation}
For homogeneous gains, $g=0$ and $\rho=\max_{i\ge2}|1-\beta-\alpha\lambda_i(L)|$, and~\eqref{eq:spectral} reduces to the classical bound $\bar d_s/(1-\rho)$.
\end{corollary}

Corollary~\ref{cor:spectral} extends the classical bound to heterogeneous gains.
For homogeneous gains, $\rho$ depends only on the eigenvalues of $L$, and a larger $\lambda_2(L)$ typically reduces $\rho$ and therefore the ideal disagreement.
With heterogeneous gains, the term $g^2/N$ reduces $\Delta$.
This is because the gain spread converts disagreement into an offset of the network average, which then re-enters the disagreement through the innovation term.
Note that we use the exact term~\eqref{eq:edis} for network design and~\eqref{eq:spectral} only to explain the roles of connectivity and gain heterogeneity.

\section{Network Design}\label{sec:netdesign}

A configuration $(\mathcal G,\mathcal B)$ is feasible if it satisfies the per-agent and network-wide bit budgets of Section~\ref{sec:problemdef} and uses only links in the set $\mathcal E_{\mathrm{feas}}$ of physically feasible links.
With innovation gains defined in~\eqref{eq:gains}, the design problem is
\begin{align}
    \nonumber \min_{\mathcal E,\{b_{nm}\}}\quad& J(\mathcal E,\mathcal B)=\min_{\alpha\in\mathcal A(L)}\etot(L,\mathcal B,\alpha)+\mu\!\!\sum_{(n,m)\in\mathcal E}\!\!b_{nm}\\
    \nonumber \mathrm{s.t.}\quad&\lambda_2(L)\ge\underline\lambda_2,\quad \textstyle\sum_mb_{nm}\le\bar B_n,\quad b_{nm}\in\mathcal B^\star\\
    &\textstyle\sum_{(n,m)\in\mathcal E}b_{nm}\le B_{\mathrm{tot}},\quad \mathcal E\subseteq\mathcal E_{\mathrm{feas}},
\label{eq:design}
\end{align}
where $\mathcal B^\star$ is the finite set of admissible link resolutions and $\mu\ge0$ is an optional rate penalty.
The constraint $\lambda_2(L)\ge\underline\lambda_2>0$ ensures that the network remains connected.
The consensus gain is selected from the admissible set $\mathcal A(L)\subset(0,\bar\alpha(L,\betab)]$, which is a logarithmic grid of $J_\alpha$ points.
Therefore, every selected gain satisfies Assumption~\ref{ass:step}.
Note that the consensus gain must be selected jointly with the network.
A larger $\alpha$ strengthens the consensus, but it also increases the quantization load $q_n$ in~\eqref{eq:loads}.
Therefore, neither the graph nor the link rates determine the best configuration alone.

Solving~\eqref{eq:design} exactly requires a search over all graphs and bit allocations, and each evaluation of $\etot$ requires $O(N^3)$ matrix operations.
Therefore, we propose the \ac{jenc} algorithm, summarized in Algorithm~\ref{alg:jenc}, which searches~\eqref{eq:design} through one of the following single-link actions: (i) adding a link at a given rate, (ii) changing the rate of a link, or (iii) removing a link.
In what follows, the details of \ac{jenc} Algorithm are explained. First, the endpoint agents of each link rank the candidate actions using local information, and a coordinator verifies the selected actions using the exact objective.
\ac{jenc} starts from the initial configuration $(\mathcal G_0,\mathcal B_0)$, verifies its feasibility and selects the consensus gain that minimizes $\etot$ over $\mathcal A(L)$ (Line~\ref{alg:init}).

\begin{algorithm}[t]
    \footnotesize
    \DontPrintSemicolon%
    \SetAlgoInsideSkip{smallskip}
    \caption{Joint Estimation and Network Configuration (JENC)}\label{alg:jenc}
    \KwIn{$(L_0,\mathcal B_0)$; $\mathcal E_{\mathrm{feas}}$; $\mathcal B^\star$; budgets; delays $\{\tau_{nm}\}$; variances $\{\sigma_n^2\}$; margin $\delta>0$; connectivity floor $\underline\lambda_2$; sweep cap $K$.}
    \KwOut{$(L,\mathcal B,\alpha)$.}
    Initialize $(L,\mathcal B)\leftarrow(L_0,\mathcal B_0)$; verify feasibility; compute $J$ and $\alpha=\arg\min_{\mathcal A(L)}\etot$\;\nllabel{alg:init}
    \For{$k=1,\ldots,K$}{
        Coordinator broadcasts $\alpha_k\leftarrow\alpha$; $\mathcal P\leftarrow\emptyset$\;\nllabel{alg:broadcast}
        \ForEach{rate action $(n,m,b)$ with $b\in\mathcal B^\star$ and each removal of an active link}{\nllabel{alg:actions}
            \If{unchanged, or violates the budgets or connectivity floor}{\nllabel{alg:filter}continue}
            Endpoints evaluate~\eqref{eq:endpoint_values}, exchange quantized bids, and form $s_{nm}$\;\nllabel{alg:bid}
            \If{$s_{nm}>\eta_{\mathrm{switch}}$}{\nllabel{alg:propose}$\mathcal P\leftarrow\mathcal P\cup\{(\text{action},s_{nm})\}$}
        }
        \ForEach{action in $\mathcal P$ in decreasing order of score}{\nllabel{alg:certify_start}
            Form $(L^{+},\mathcal B^{+})$; \textbf{if} feasible, $J^{+}\le J-\delta$, and $\etot^{+}\le\etot$\nllabel{alg:bound}\;
            \quad\textbf{then} accept and update $(L,\mathcal B,J,\alpha)$ \textbf{else} discard\;
        }\nllabel{alg:certify_end}
        \If{no action was accepted in this sweep}{\textbf{terminate}: no acceptable score-qualified action}\nllabel{alg:converge}
    }
    Repeat Lines~\ref{alg:init}--\ref{alg:converge}, first with rate actions on active links only; return the pass with the lower $J$\;\nllabel{alg:pass}
\end{algorithm}

At the beginning of each sweep, the coordinator broadcasts the current consensus gain $\alpha_k$ (Line~\ref{alg:broadcast}).
For each candidate action on link $(n,m)$ with the candidate state $(A^{+},\mathcal B^{+})$ (Line~\ref{alg:actions}), each endpoint $i\in\{n,m\}$ computes its local value
\begin{align}
    \nonumber g_i=&\;w_d\big(D_i(A,\mathcal B)-D_i(A^{+},\mathcal B^{+})\big)+w_f\big(F_i(A)-F_i(A^{+})\big)\\
    &+w_c\big(R_i(A^{+})-R_i(A)\big)-w_b\big(U_i(\mathcal B^{+})-U_i(\mathcal B)\big)
\label{eq:endpoint_values}
\end{align}
from one- and two-hop information, where $w_d$, $w_f$, $w_c$, and $w_b$ are nonnegative weights.
For the removal of a link, $g_i$ is computed in the same way as the candidate state $(A^{-},\mathcal B^{-})$.
The first two terms in~\eqref{eq:endpoint_values} use
\begin{equation}
    D_i=\frac{\alpha_k\sum_ja_{ij}c\,2^{-b_{ij}}}{\beta_i+\alpha_kd_i},\qquad
    F_i=\Big(\sigma_i^{-2}+\textstyle\sum_{j\in\mathcal N_i}\sigma_j^{-2}\Big)^{-1/2},
    \label{eq:local_terms}
\end{equation}
where $D_i$ approximates the communication bound of Agent~$i$, and $F_i$ is the per-coordinate \ac{blue} error of the measurements in the neighborhood of Agent~$i$.
Note that $D_i$ is a one-step Jacobi approximation of~\eqref{eq:xbar_wc}, since $\xbar$ satisfies $(\beta_i+\alpha d_i)\xbar_i=q_i+\alpha\sum_ja_{ij}\xbar_j$.
The third term rewards reachability through $R_i(A)=|\mathcal N_i\cup\mathcal N_i^{(2)}|-w_o\sum_{j\in\mathcal N_i}|\mathcal N_i\cap\mathcal N_j|$, which counts the one- and two-hop neighbors with an overlap discount, where $\mathcal N_i^{(2)}=\bigcup_{j\in\mathcal N_i}\mathcal N_j\setminus\{i\}$.
The last term penalizes the local bit usage $U_i=\sum_jb_{ij}$.
None of these terms is the exact reduction of $\etot$ caused by the action. They are local proxies for it.

Then, the endpoints exchange their quantized bids $\hat g_n$ and $\hat g_m$ over the link or a control channel using $b_{\mathrm{msg}}$ bits, and they form the bilateral score (Line~\ref{alg:bid})
\begin{equation}
\begin{aligned}
    s_{nm}&=\hat g_n+\hat g_m-2\eta_{nm}-w_{\mathrm{sw}},\\
    \eta_{nm}&=w_{\mathrm{msg}}c_{\mathrm{msg}}2^{-b_{\mathrm{msg}}}+w_\tau\tau_{nm},
\end{aligned}
    \label{eq:update_score}
\end{equation}
where the margin $\eta_{nm}$ accounts for the bid quantization and the bid-exchange delay, and $w_{\mathrm{sw}}$ is a switching cost.
A proposed action is sent to the coordinator by an agent only if it changes the configuration, satisfies the endpoint budgets and the connectivity floor (Line~\ref{alg:filter}), and has a score $s_{nm}>\eta_{\mathrm{switch}}$ (Line~\ref{alg:propose}).

Next, the coordinator processes the proposals in decreasing order of score (Lines~\ref{alg:certify_start}--\ref{alg:certify_end}).
For each proposal, it recomputes the feasibility and the objective in the current configuration.
The action is accepted only if $J(\mathcal E^{+},\mathcal B^{+})\le J(\mathcal E,\mathcal B)-\delta$ for a margin $\delta>0$, and $\etot(\mathcal E^{+},\mathcal B^{+})\le\etot(\mathcal E,\mathcal B)$ (Line~\ref{alg:bound}).
The second test ensures that the bound does not increase when a rate penalty is used and is redundant for $\mu=0$.
\ac{jenc} terminates when no action is accepted in a sweep (Line~\ref{alg:converge}), or after the sweep cap $K$.
Therefore, \ac{jenc} is semi-distributed: the endpoint agents compute the bids, whereas the coordinator evaluates the global matrices and verifies the feasibility and descent.
Note that an accepted shortcut increases the degrees and therefore reduces the admissible gain $\bar\alpha$, and no single action can undo it.
Therefore, the coordinator runs a second pass, called the rate-first pass, from $(\mathcal G_0,\mathcal B_0)$ that first proposes only rate actions on the active links and returns the pass with the lower $J$ (Line~\ref{alg:pass}).

\begin{proposition}[Properties of JENC]\label{prop:auction}
Algorithm~\ref{alg:jenc} satisfies the following properties.
\begin{itemize}
    \item[(a)] Every iteration is feasible with a connected graph, and every deployed consensus gain satisfies Assumption~\ref{ass:step}.
    \item[(b)] Each accepted action decreases $J$ by at least $\delta$; hence, without the sweep cap, each pass performs at most $\lceil(J_0-\epsilon_{\mathrm{fus}})/\delta\rceil$ accepted actions and cannot cycle.
\item[(c)] Upon termination of a pass after a sweep without an accepted action, no feasible, score-qualified single-link action satisfies both the $\delta$-descent test and the nonincreasing-bound test. For $\mu=0$, no such action decreases $J$ by at least $\delta$.
\end{itemize}
\end{proposition}

Proposition~\ref{prop:auction} shows that every accepted action reduces $J$ and \ac{jenc} Algorithm converges.

\emph{Complexity:} Let $M_c$ denote the number of feasible links, $L_b=|\mathcal B^\star|$, and $P$ the number of proposals in a sweep.
Each endpoint computes its bids with $O(d_cL_b)$ operations, where $d_c$ is the candidate degree, and sorting all actions requires $O(M_cL_b\log(M_cL_b))$ operations.
The coordinator evaluates the bound once per proposal at a cost of $O(J_\alpha N^3)$, i.e., $O(PJ_\alpha N^3)$ per sweep.
In comparison, a centralized greedy configurator evaluates every feasible action, at a cost of $O(M_cL_bJ_\alpha N^3)$ per accepted action.

\section{Numerical Evaluation}\label{sec:evaluation}
\begin{figure}[b]
    \centering
    \includegraphics[width=0.82\linewidth]{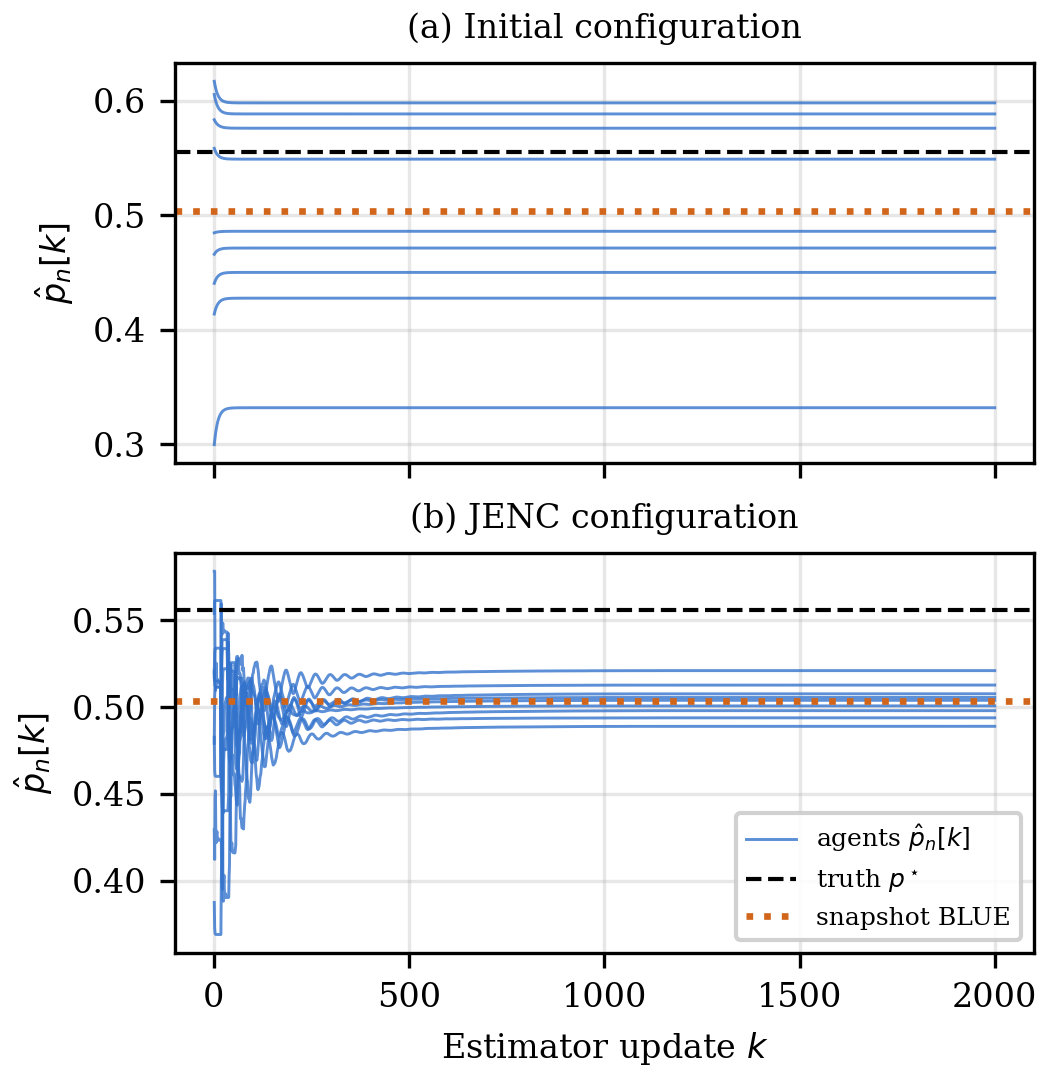}
    \caption{Estimates of a 9-agent MAS for one measurement snapshot under (a) the initial configuration (line, 2-bit links) and (b) the JENC configuration.}
    \label{fig:illustrative}
\end{figure}

\begin{figure*}[t]
    \centering
    \subfloat[Disagreement]{\includegraphics[width=0.3\linewidth]{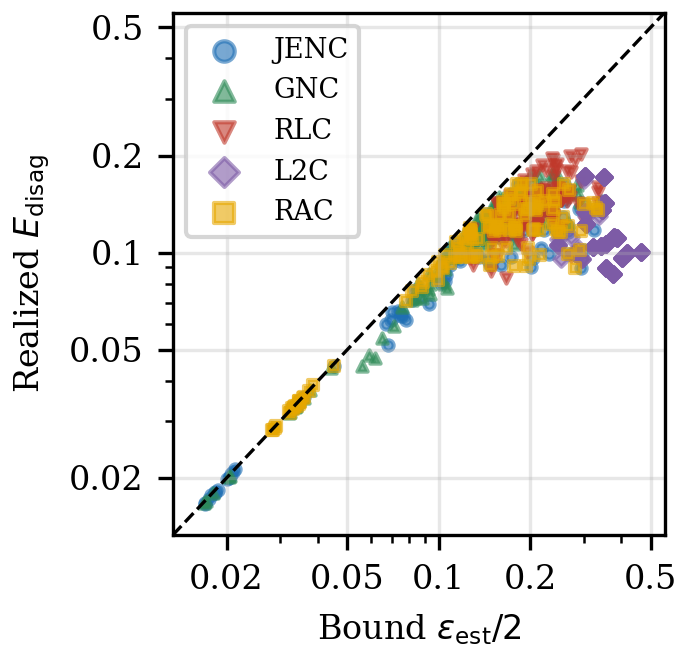}\label{fig:validity_disag}}
    \hfil
    \subfloat[Inaccuracy]{\includegraphics[width=0.3\linewidth]{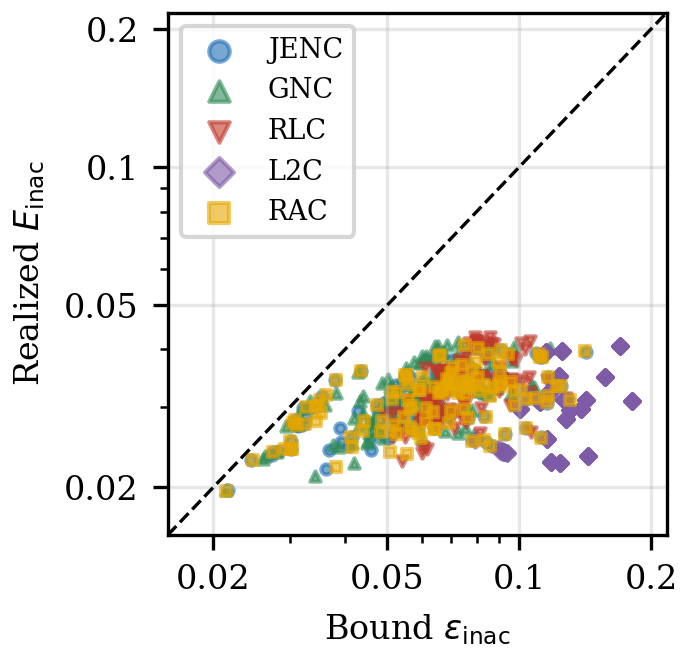}\label{fig:validity_inac}}
    \hfil
    \subfloat[Total error]{\includegraphics[width=0.3\linewidth]{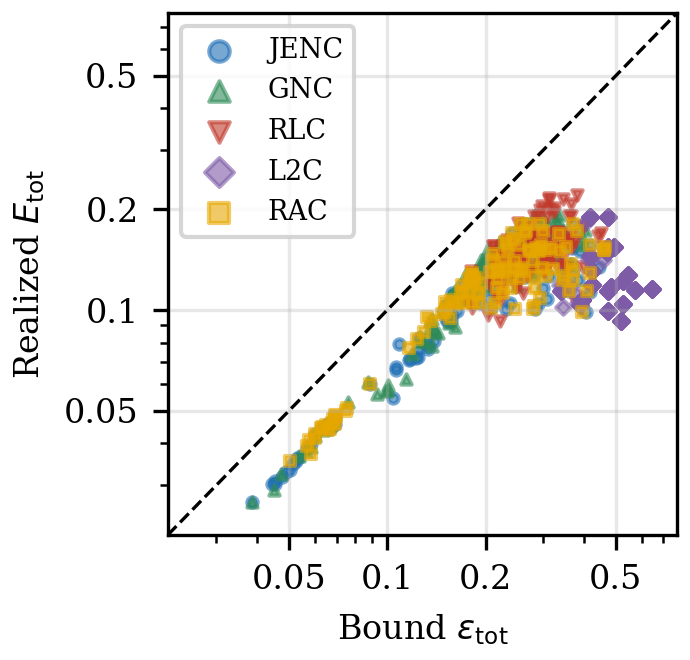}\label{fig:validity_tot}}
    \caption{Bound vs. realized error ($5$ methods, $20$ instances, $6$ budgets, $8192$ snapshots each). Points below the dashed diagonal satisfy the bound.}
    \label{fig:validity}
\end{figure*}

\begin{figure*}[t]
    \centering
    \subfloat[Realized total error]{\includegraphics[width=0.3\linewidth]{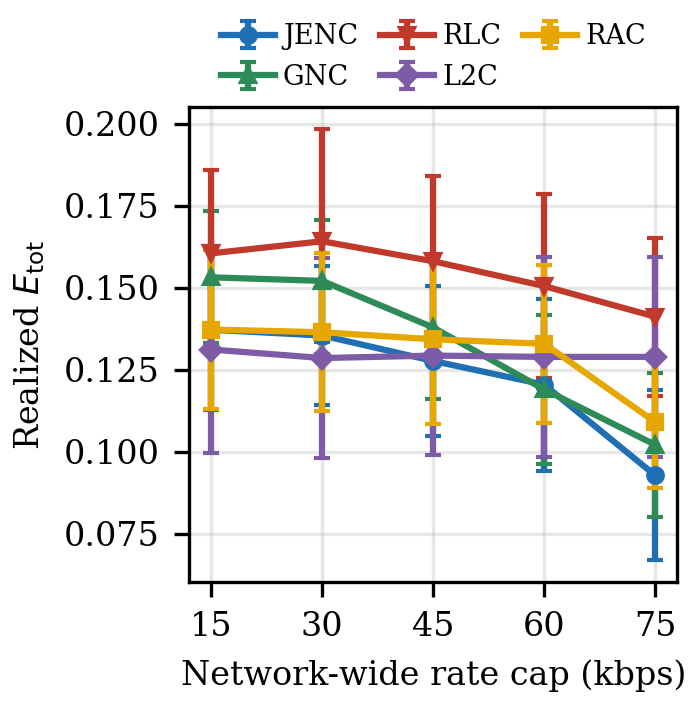}\label{fig:native_error}}
    \hfil
    \subfloat[Total-error bound]{\includegraphics[width=0.3\linewidth]{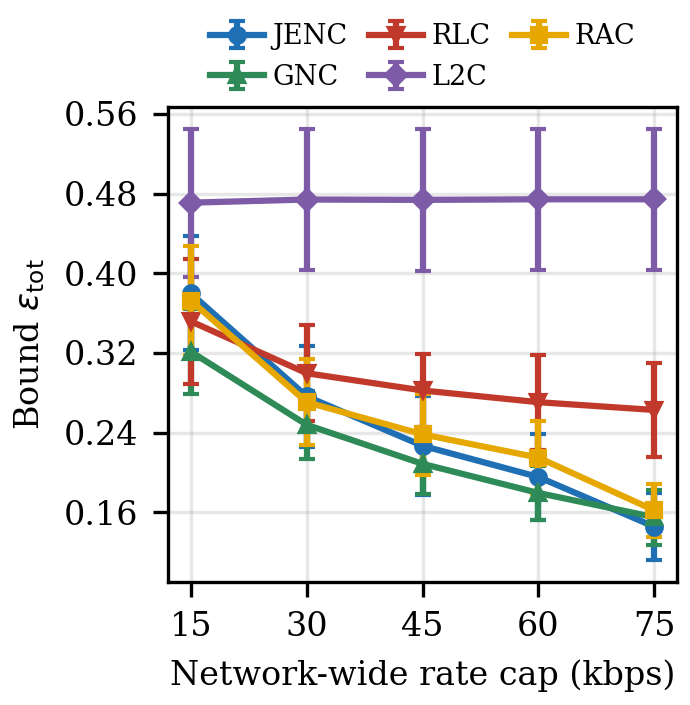}\label{fig:native_bound}}
    \hfil
    \subfloat[Algebraic connectivity]{\includegraphics[width=0.3\linewidth]{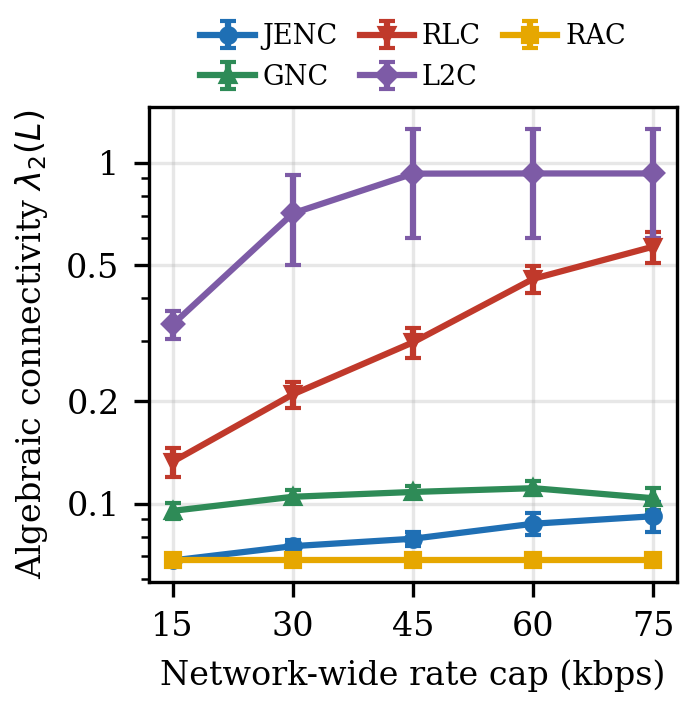}\label{fig:native_lambda2}}
    \caption{Comparison under matched network-wide rate budgets over $20$ instances. Each method operates at its selected rates and satisfies its budget.}
    \label{fig:native}
\end{figure*}

We implemented the estimator, the bound in Theorem~\ref{thm:total}, \ac{jenc}, and four baselines in Python.
The baselines are (i) GNC, a centralized greedy configurator that selects the feasible single-link action with the largest exact decrease in $J$, (ii) RLC, an action-value heuristic driven by the same objective, (iii) L2C, a spectral configurator that greedily adds the feasible minimum-rate link with the largest increase in $\lambda_2$~\cite{ghosh2006growing,kar2008randomlinks}, and (iv) RAC, a rate-only configurator that applies the same exact greedy descent as GNC to the link rates of the initial graph without changing the topology.
All methods share the same initial configuration and assumptions. 

Each test case consists of $N=12$ agents unless otherwise mentioned.
The initial graph is a line and the delay of each link is a random number between 14 and 22.
Sensor standard deviations are drawn log-uniformly from $[0.05,0.30]$, and the scalar state is drawn uniformly from $[0.4,0.6]$.
The link rates $\{1,2,4,8\}$~kbps correspond to $\{2,4,8,16\}$ bits per update at $T_s=2$~ms.
We use a uniform deterministic quantizer that satisfies~\eqref{eq:quantization} with $c=1$, i.e., $\kappa=1$ around the center $0.5$ and
spacing $2\kappa 2^{-b}$.
Since the simulated noise is Gaussian, the measurements are not bounded; therefore, we simulate the quantizer without a range limit and verify that at most $0.0164\%$ of its inputs leave the range $[-0.5,1.5]$ in any configuration.
The remaining parameters are listed in Table~\ref{tab:params}.
Each configuration is simulated for 4000 updates with 256 independent measurement snapshots.
The expectations in~\eqref{eq:Etot}, \eqref{eq:Eagr}, and~\eqref{eq:Eacc} are estimated over the snapshots and the final 1000 updates. We refer to these estimates as realized errors.

\begin{table}[t]
    \centering
    \footnotesize
    \caption{Simulation parameters}
    \label{tab:params}
    \resizebox{\columnwidth}{!}{%
    \begin{tabular}{ll}
        \toprule
        Parameter & Value \\
        \midrule
        Innovation scale $\beta_0$; profile & $0.2$; inverse variance~\eqref{eq:gains} \\
        Consensus gain $\alpha$ & $\arg\min_{\mathcal A(L)}\etot$, $J_\alpha=48$ points, log-spaced on $[\bar\alpha/50,\bar\alpha]$ \\
        Quantizer constant $c$; bid constant $c_{\mathrm{msg}}$ & $1.0$; $0.5$ \\
        Update period $T_s$; rate set $\mathcal B^\star$ & $2$~ms; $\{1,2,4,8\}$~kbps ($\{2,4,8,16\}$ bits/update) \\
        Per-agent budget $\bar B_n$; network budgets & $18$~kbps; $\{15,30,45,60,75\}$~kbps and uncapped \\
        Rate penalty $\mu$; descent margin $\delta$ & $0.01$; $10^{-10}$ \\
        Weights $w_d,w_f,w_c,w_b$ & $100$, $5$, $0.3$, $5\times10^{-4}$ \\
        Weights $w_{\mathrm{sw}},w_{\mathrm{msg}},w_\tau,w_o$; threshold $\eta_{\mathrm{switch}}$ & $0.1$, $1.0$, $0.02$, $0.25$; $0.15$ \\
        $\lambda_2$ floor $\underline\lambda_2$ & $0.95\lambda_2(L_0)$ \\
        Sweep or round cap $K$ & $60$ (all methods) \\
        Instances (main study / sweeps) & $20$ / $10$ \\
        \bottomrule
    \end{tabular}}
\end{table}

Fig.~\ref{fig:illustrative} shows the estimates of a 9-agent \ac{mas} for one snapshot of measurement.
In the initial configuration, agents communicate on a graph with $\lambda_2 = 0.121$ with 2-bit links.
The network graph, the link rates, and the gains are not optimized to reduce the estimation error.
As a result, the agents do not agree and their estimates remain spread over a wide range.
On the other hand, \ac{jenc} uses the gains in~\eqref{eq:gains} and optimizes the network configuration.
This increases $\lambda_2$ to $0.468$, and increases the link rates to a total of $72$~kbps.
Therefore, the estimates converge close to the snapshot \ac{blue} and the bound decreases from $0.374$ to $0.047$, i.e., by $87.4\%$.

Fig.~\ref{fig:validity} compares the bounds with the realized errors of all $600$ configured networks, which we re-estimate with $8192$ snapshots for this comparison.
Every realized total estimation error satisfies its bound and the ratio between the bounds and  the realized errors ranges from $0.18$ to $0.75$, with a median of $0.56$, see Fig.~\ref{fig:validity_tot}.
Therefore, the bound is valid but conservative.
The inaccuracy bound is also satisfied in every configuration, with a median ratio of $0.46$ (Fig.~\ref{fig:validity_inac}).
The disagreement bound is tighter, with a median ratio of $0.70$ (Fig.~\ref{fig:validity_disag}).
Its estimate is also satisfied in all configurations.

However, as shown in the figure, the bound ranks the configurations only weakly. Its correlation with the total realized error is $0.503$, and the smallest bound identifies the smallest realized error in $51$ of the $120$ comparisons.
This is because Lemma~\ref{lem:wc} assumes that quantization errors accumulate in the worst case, whereas realized errors partially cancel across agents and updates.


\begin{table}[t]
    \centering
    \footnotesize
    \caption{Mean realized $\Etot$ over $20$ instances. Bold: lowest mean. In parentheses: number of instances where JENC achieves a lower error. The initial configuration achieves $\Etot=0.135$.}
    \label{tab:native}
    \resizebox{\columnwidth}{!}{%
    \begin{tabular}{lcccccc}
        \toprule
        Method & 15~kbps & 30~kbps & 45~kbps & 60~kbps & 75~kbps & Uncapped \\
        \midrule
        JENC & $0.133$ & $0.131$ & $\mathbf{0.128}$ & $\mathbf{0.120}$ & $\mathbf{0.093}$ & $\mathbf{0.039}$ \\
        GNC & $0.153\,(19)$ & $0.152\,(19)$ & $0.138\,(17)$ & $0.121\,(10)$ & $0.102\,(13)$ & $0.052\,(14)$ \\
        RLC & $0.160\,(18)$ & $0.164\,(16)$ & $0.158\,(20)$ & $0.150\,(18)$ & $0.141\,(20)$ & $0.139\,(20)$ \\
        L2C~\cite{ghosh2006growing,kar2008randomlinks} & $\mathbf{0.130}\,(9)$ & $\mathbf{0.129}\,(9)$ & $0.129\,(11)$ & $0.129\,(13)$ & $0.129\,(18)$ & $0.129\,(20)$ \\
        RAC (rate only) & $0.137\,(9)$ & $0.136\,(10)$ & $0.134\,(11)$ & $0.133\,(15)$ & $0.109\,(14)$ & $0.046\,(13)$ \\
        \bottomrule
    \end{tabular}}
\end{table}

Fig.~\ref{fig:native} and Table~\ref{tab:native} compare \ac{jenc} with the baselines under network-wide budgets of 15--75~kbps and without a network-wide cap.
The results show three advantages of \ac{jenc}.
First, averaged over the capped budgets, \ac{jenc}'s realized error is $0.123$, compared with $0.130$ for RAC, $0.129$ for L2C, $0.133$ for GNC, and $0.155$ for RLC. Without a network-wide cap, \ac{jenc} achieves the lowest error of $0.039$.
Fig.~\ref{fig:native_lambda2} shows that \ac{jenc} keeps $\lambda_2$ close to that of the initial line ($0.08$ versus $0.07$), while L2C and RLC raise it to $0.77$ and $0.33$, respectively, and still achieve larger errors.
Therefore, \ac{jenc} reduces the error more than the baselines by jointly allocating the link rates and the consensus gain.
Note that L2C achieves a lower mean estimation error at 15 and 30~kbps, which is very close to \ac{jenc}.
However, it uses $16.0$ links instead of $11.1$, and its bound remains close to $0.471$.
Additionally, at 15~kbps, \ac{jenc} outperforms GNC in $19$ of $20$ instances.
Despite this, its bound is higher because the worst-case communication term dominates the bound under tight budgets.

\begin{figure}[t]
    \centering
    \includegraphics[width=0.9\linewidth]{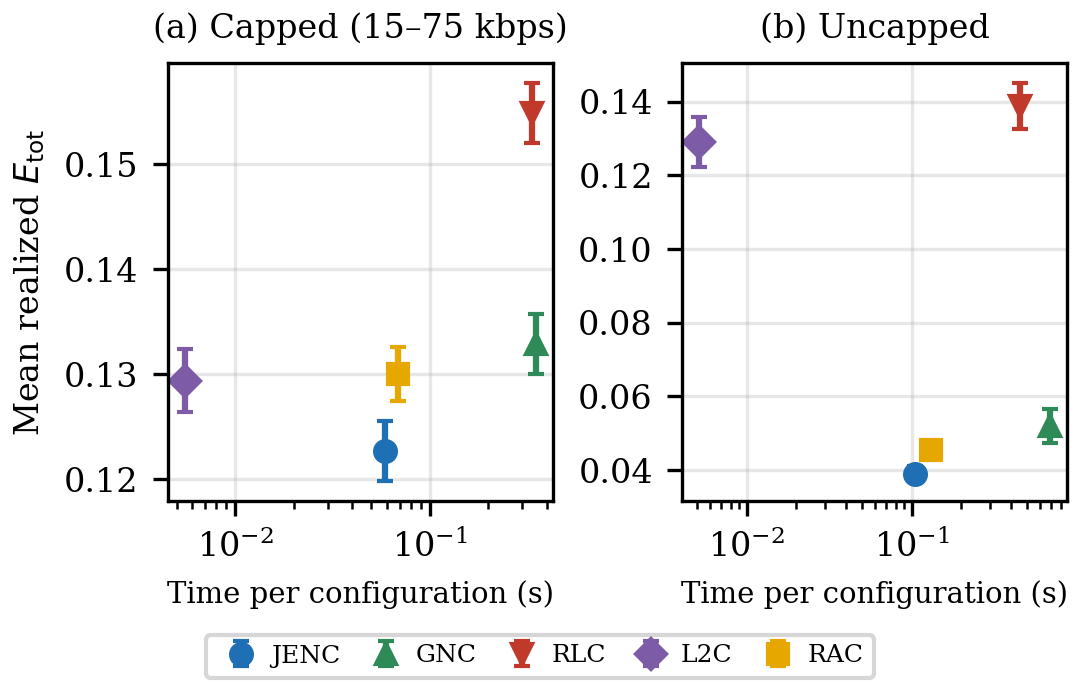}
    \caption{Mean realized total error versus the configuration time of each method, (a) over the capped budgets and (b) without a network-wide cap (mean $\pm$ one standard error).}
    \label{fig:quality_cost}
\end{figure}

Second, as shown in Fig.~\ref{fig:quality_cost}, \ac{jenc} requires less computation than the other bound-driven methods. Its runtime is $0.07$~s per configuration, compared with $0.08$~s for RAC, $0.36$~s for RLC, and $0.41$~s for GNC.
This is because \ac{jenc} evaluates the bound only for the proposals that pass the endpoint scores, i.e., $52$ times per configuration, compared with $1045$ times for GNC.
For $N=50$ agents, \ac{jenc} requires $1.0$~s, while GNC requires $35.3$~s.
Additionally, the endpoints exchange only $551$ quantized bids, i.e., about $0.3$~kB, 
and the coordinator collects the variances and the feasible links only once.

Third, \ac{jenc} outperforms the baselines that do not use the bound effectively.
Without a network-wide cap, it outperforms L2C and RLC in all instances, because L2C retains 2-bit links and RLC adds many links as discussed earlier.

Finally, we run a ratio consensus, see Section~\ref{sec:innogains}, on \ac{jenc} configurations with $b$ bits for each of its two values, which favors the baseline.
Over lossless links, it achieves the error provided by \ac{blue} ($0.0278$).

However, in capped quantized links, quantization error accumulates in the network sum of the weights~$\sigma_n^{-2}$, which drifts by $25\%$.
As a result, the mean error of the ratio consensus is $0.439$, compared with $0.123$ for \ac{jenc}.
With $b/2$ bits for each value, i.e., the same number of bits per link as \ac{jenc}, the error of ratio consensus increases to $0.509$.
Without a network-wide cap where most links use 16 bits, ratio consensus achieves a lower error ($0.028$) compared with \ac{jenc}.
Therefore, the innovation term is advantageous when link rates are limited.


Table~\ref{tab:overhead} reports the number of active links, the messages per estimator update, the topology changes relative to the initial line, and the bound evaluations during configuration.
Since each link has its own bit rate, messages are sent over dedicated (unicast) links, so the number of links determines the number of transmissions.
L2C and RLC add many links. They use $16.0$ and $14.7$ links on average, i.e., $32.0$ and $29.5$ messages per update, compared with $22.1$ for \ac{jenc}.
These additional transmissions increase the channel load and interference. The resulting larger $\lambda_2$ (Fig.~\ref{fig:native_lambda2}) does not reduce their realized errors below that of \ac{jenc}.
Additionally, on average, \ac{jenc} changes the topology only $0.7$ times per configuration compared with $4.0$ for GNC, $5.0$ for L2C, and $8.7$ for RLC.
Therefore, \ac{jenc} reduces the error first by reallocating the bit rates of the existing links and then by changing the network topology.
Note that the bound in~\eqref{eq:etot} is a steady-state bound, and therefore the design does not account for the convergence time.
Over the capped budgets, as shown in Table~\ref{tab:overhead}, all methods settle within a similar number of updates.
However, without a network-wide cap, \ac{jenc} and RAC settle in $688$ and $746$ updates, respectively, compared with $415$ for GNC.
Therefore, under loose budgets, the lower steady-state error of \ac{jenc} comes at the cost of a longer transient.

\begin{figure}[t]
    \centering
    \includegraphics[width=0.9\linewidth]{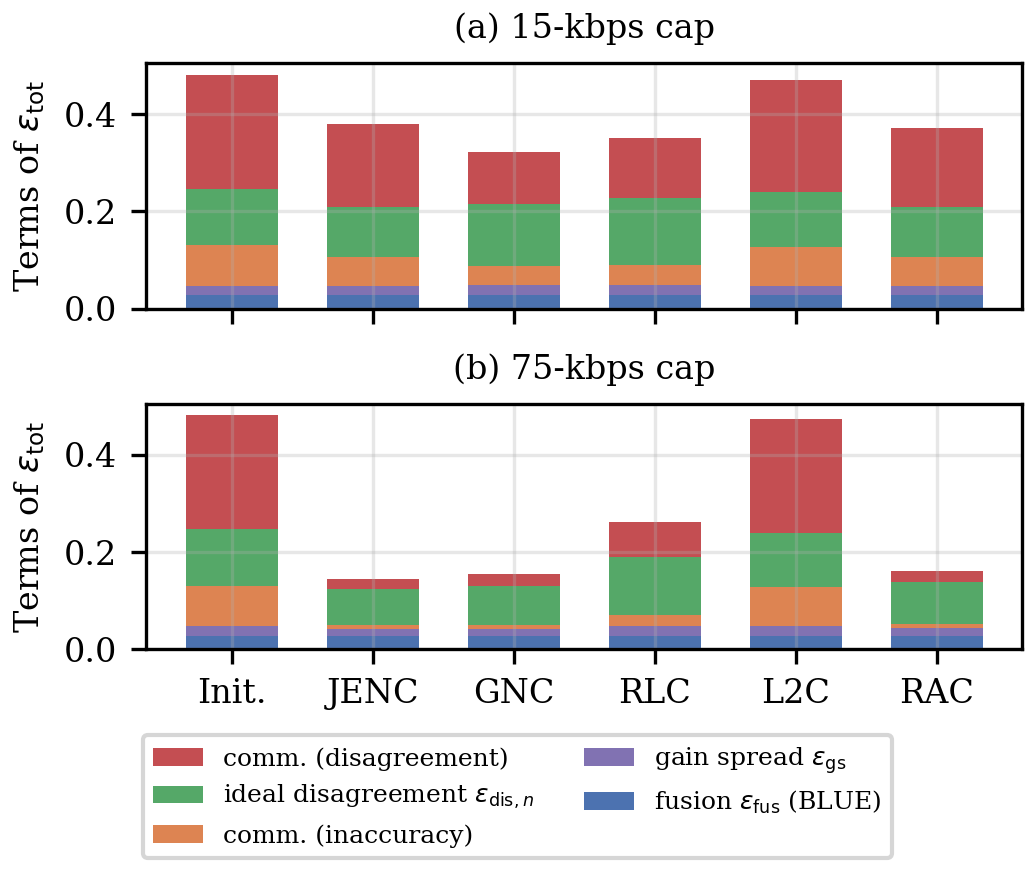}
    \caption{Mean components of $\etot$ at (a) the tightest and (b) the loosest capped budget.}
    \label{fig:decomposition}
\end{figure}

Fig.~\ref{fig:decomposition} shows the components of the bound.
The fusion term is equal to the error of \ac{blue} which is approximately $0.0273$ and does not depend on the configuration.
At 15~kbps, the communication term is the largest component and the bound selects a small consensus gain, $\alpha\approx0.008$, because $q_n$ is proportional to $\alpha$.
As the budget increases, higher link resolutions reduce the communication term, and the selected gain increases to $\alpha\approx0.16$ without a cap.
Therefore, the ideal disagreement of \ac{jenc} decreases by more than half, and the gain-spread term decreases with it.
This is consistent with~\eqref{eq:gain_spread}.

\begin{table}[t]
    \centering
    \footnotesize
    \caption{Configuration and communication cost (mean over the capped budgets). L2C does not use the bound.}
    \label{tab:overhead}
    \resizebox{\columnwidth}{!}{%
    \begin{tabular}{lcccccc}
        \toprule
        Method & Links & Messages & Topology changes & Bound evaluations & Time (s) & Settling \\
        \midrule
        JENC & $11.1$ & $22.1$ & $0.7$ & $52$ & $0.06$ & $208$ \\
        GNC & $11.2$ & $22.3$ & $4.0$ & $1045$ & $0.35$ & $208$ \\
        RLC & $14.7$ & $29.5$ & $8.7$ & $111$ & $0.34$ & $230$ \\
        L2C & $16.0$ & $32.0$ & $5.0$ & -- & $0.01$ & $209$ \\
        RAC & $11.0$ & $22.0$ & $0.0$ & $210$ & $0.07$ & $219$ \\
        \bottomrule
    \end{tabular}}
\end{table}

\begin{figure}[t]
    \centering
    \includegraphics[width=0.9\linewidth]{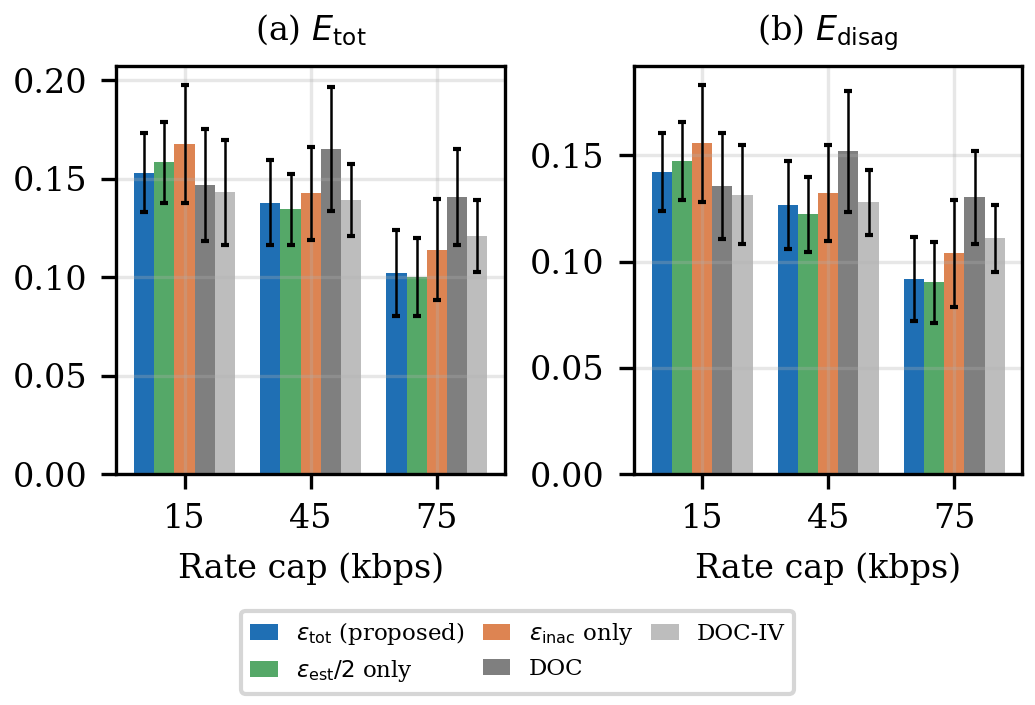}
    \caption{Realized (a) total error and (b) disagreement over 20 instances. All configurations are obtained by GNC and deployed with the same estimator, with $\alpha$ selected by $\etot$.}
    \label{fig:objective}
\end{figure}

The previous comparisons fix the design objective to $\etot$ for \ac{jenc} and the baselines.
Here, we use different objectives for a single method, i.e. GNC, to show whether $\etot$ is a good design cost objective and how it compares with other objectives.
Fig.~\ref{fig:objective} compares five design objectives: the total-error bound $\etot$, the disagreement bound $\tfrac12\eest$, the inaccuracy bound $\eacc$, a disagreement-only bound (DOC) that represents a disagreement-centered design~\cite{olfati2004time,kar2010quantized,chen2019bit}, and DOC with gains in \eqref{eq:gains} that we name DOC-IV.
DOC uses a contraction-based disagreement bound with a staleness-based delay term and assumes homogeneous gains during design.
DOC-IV uses the same bound but with the gains defined in \eqref{eq:gains}, which are also used by the estimator in~\eqref{eq:estimator}.
All configurations are obtained by GNC and evaluated with the same estimator.

The total-error design and the disagreement-only design perform comparably at all budgets.
This is expected from Proposition~\ref{prop:structure}.
The fusion term $\epsilon_{\mathrm{fus}}$ does not depend on the configuration, the gain-spread term is bounded by the ideal disagreement in~\eqref{eq:gain_spread}, and the communication term of $\eacc$ is also contained in $\tfrac12\eest$.
Therefore, the part of the inaccuracy bound that the configuration can change is already controlled by the disagreement bound.
To verify this, we repeated the comparison on the held-out instances with noise scales up to 4 and up to twice the nominal heterogeneity of $\sigma_n$, i.e., in $18$ regimes.
The total-error design outperforms the disagreement-only design in less than $4$ of $6$ instances.
Note that the inaccuracy bound still matters for the estimator.
It selects the gains and it bounds the accuracy, as shown in Fig.~\ref{fig:validity}.
Additionally, ignoring the disagreement increases the error. 
Relative to the inaccuracy-only design, the total-error design reduces the realized error by $7.4\%$ at 15~kbps.

Relative to DOC, the total-error design reduces the realized error by $15.4\%$ and $26.4\%$ at 45 and 75~kbps, and outperforms DOC in $19$ and $18$ of 20 instances, respectively.
DOC-IV differs from DOC only in the gains. It differs from the total-error design only in the bound.
Therefore, it separates this gap into two parts.
First, using inverse-variance gains (as in DOC) instead of homogeneous gains (as in DOC-IV) reduces the error by $13.7\%$ and $13.2\%$ at 45 and 75~kbps, respectively.
Second, using the total-error bounds instead of DOC bound,
reduces the error
by $14.9\%$ at 75~kbps, and
the total-error design outperforms
DOC-IV in $17$ of 20 instances.
Note that at 45~kbps, the difference between DOC-IV and the total-error design is not significant.
This is consistent with the delay-independent communication bound of Lemma~\ref{lem:wc}, whereas DOC charges every delayed link a staleness term.
Therefore, the benefit of the proposed analysis for the network design comes from the disagreement bound, i.e., its delay-independent communication term and its gain model, in contrast, the inaccuracy part does not change the selected configuration.
Moreover, DOC and DOC-IV add many links, i.e., respectively, $17.2$ and $19.2$ at 75~kbps, compared with $11.3$ for the total-error design.
At 15~kbps, however, the total-error design outperforms DOC-IV in only $4$ of 20 instances, consistent with the conservatism of the worst-case communication bound under tight budgets, as shown in Fig.~\ref{fig:validity}.

\subsection{Effects of Innovation Gains}\label{sec:gains}

\begin{figure}[t]
    \centering
    \subfloat[Error due to noise level]{\includegraphics[width=0.95\linewidth]{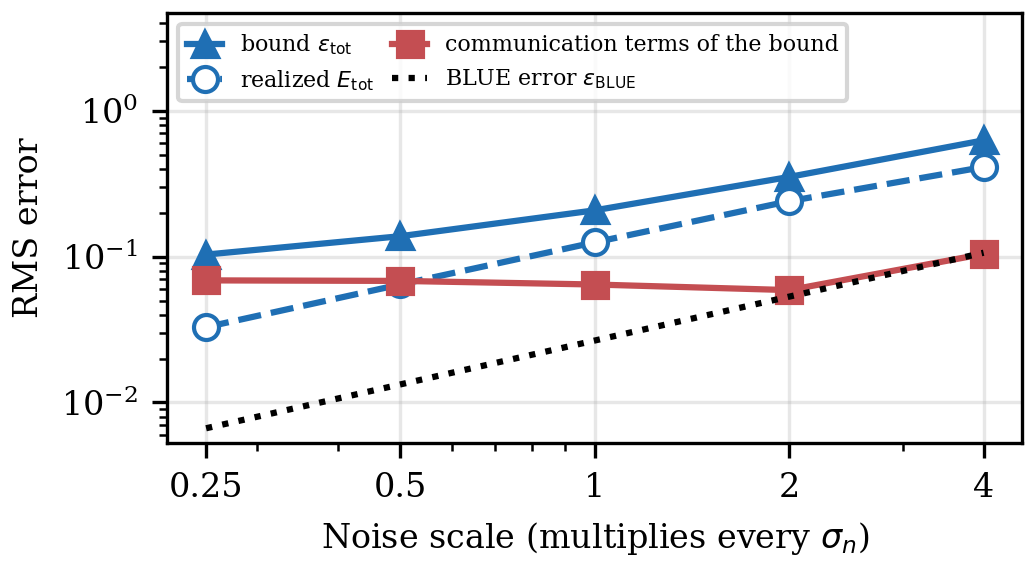}\label{fig:uncertainty_noise}}
    \\
    \subfloat[Error due to noise heterogeneity]{\includegraphics[width=0.95\linewidth]{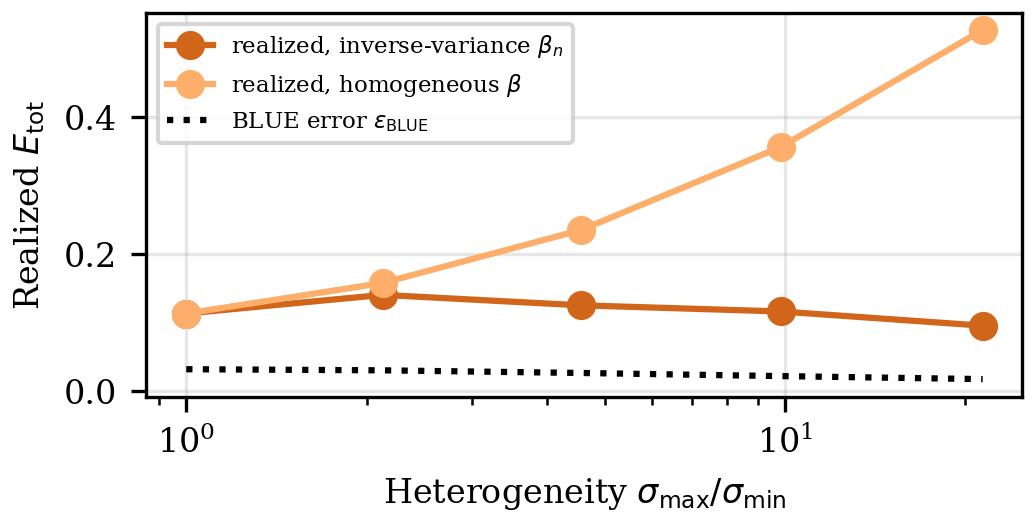}\label{fig:uncertainty_spread}}
    \caption{Effect of sensing uncertainty on JENC configurations at 45~kbps (10 instances).}
    \label{fig:uncertainty}
\end{figure}

We first compare inverse-variance gains with homogeneous gains of equal mean, each with its own \ac{jenc} configuration.
Fig.~\ref{fig:uncertainty_spread} shows the same comparison over a range of sensor heterogeneity.
At 15, 45, and 75~kbps, inverse-variance gains reduce the realized total error by $35.7\%$, $39.8\%$, and $50.9\%$, respectively, and outperform homogeneous gains in all instances.
At 45~kbps, their inaccuracy is $0.0322$, close to the \ac{blue} floor of $0.0273$, whereas homogeneous gains have a larger fusion floor of $0.0407$, as predicted by Proposition~\ref{prop:structure}(a).

\begin{figure}[t]
    \centering
    \includegraphics[width=0.9\linewidth]{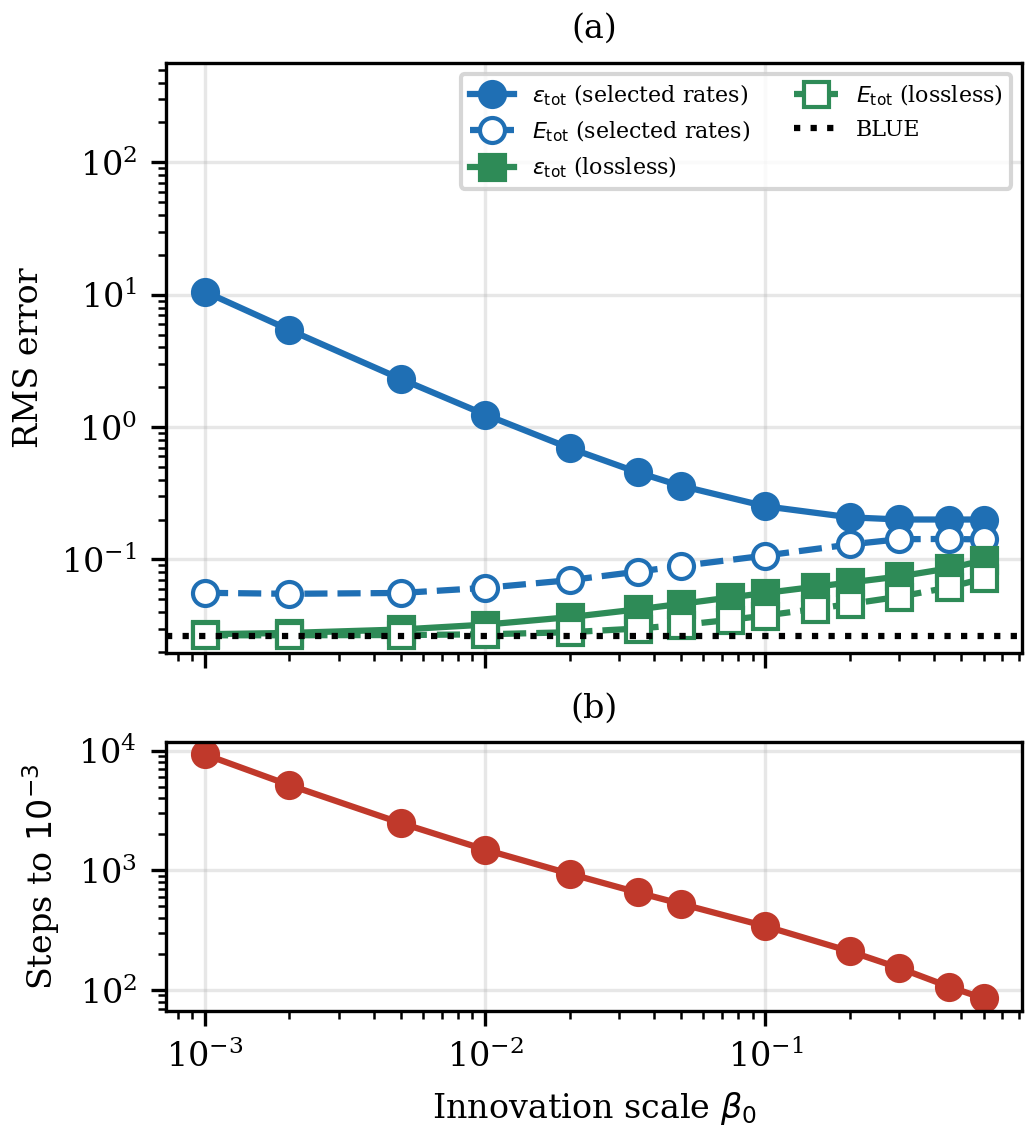}
    \caption{Effect of the innovation scale $\beta_0$ for fixed 45-kbps JENC configurations (10 instances). (a) Bound and realized error. (b) Settling time of the lossless estimator.}
    \label{fig:gains}
\end{figure}

Fig.~\ref{fig:gains} examines the innovation scale $\beta_0$.
Over lossless links, as $\beta_0$ decreases, the bound and the realized error approach the error of \ac{blue}, which is $0.0266$, as predicted by Proposition~\ref{prop:scale}(a).
Over quantized links, the bound grows to $10.528$ at $\beta_0=0.001$, as predicted by Proposition~\ref{prop:scale}(b), while the realized error decreases to $0.056$, because the realized quantization errors are partially canceled.
However, the settling time grows from $210$ updates at $\beta_0=0.2$ to $9316$ at $\beta_0=0.001$, consistent with~\eqref{eq:transient}.
Therefore, we use $\beta_0=0.2$ as a compromise between the estimation error and the convergence time.

\begin{figure}[t]
    \centering
    \subfloat[Effect of quantization on the realized error (uncapped).]{\includegraphics[width=0.85\linewidth]{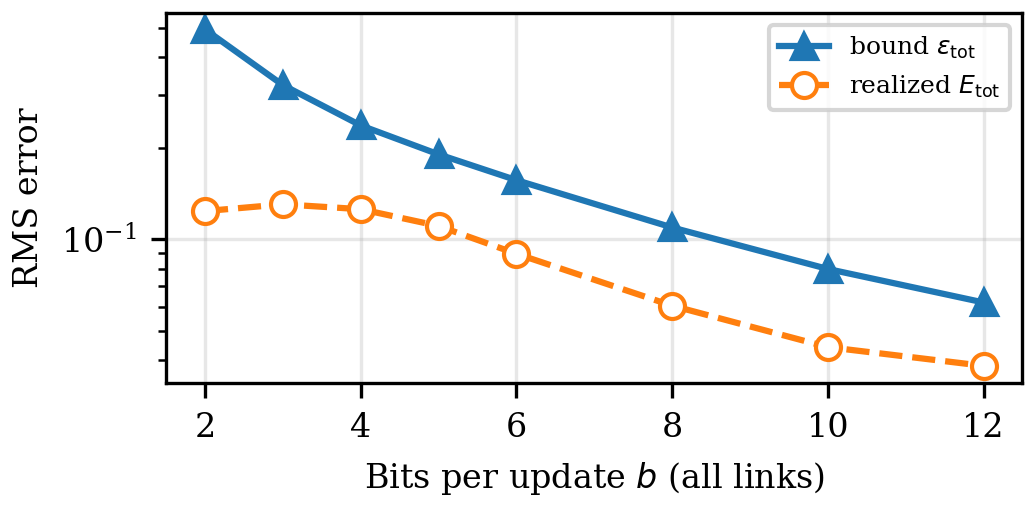}\label{fig:quant_error}}
    \\
    \subfloat[Effect of quantization on communication bound of Lemma~\ref{lem:wc}]{\includegraphics[width=0.85\linewidth]{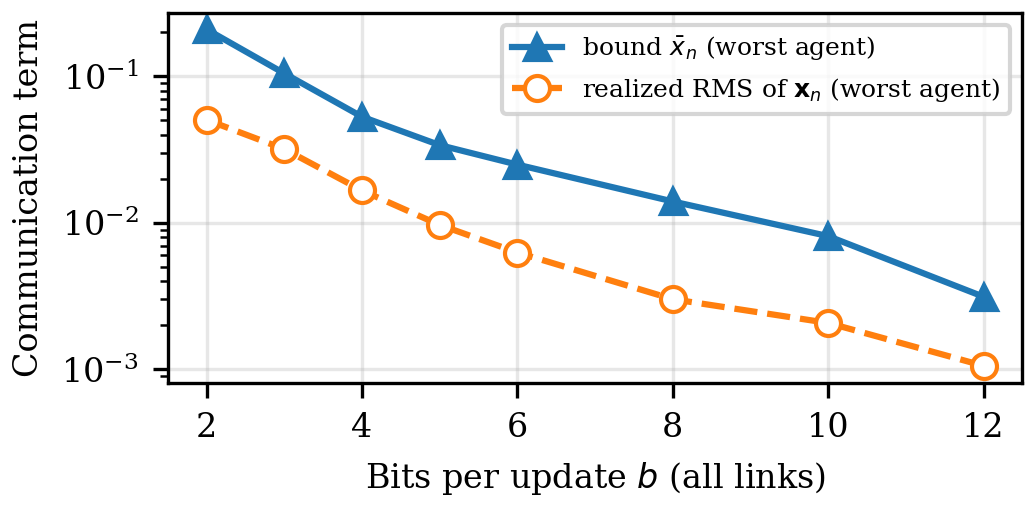}\label{fig:quant_comm}}
    \\
    \subfloat[Effect of delay on the error for 45-kbps configurations with all delays scaled: Bound (triangles), realized error (circles), and settling time of the lossless estimator (squares, right axis)]{\includegraphics[width=0.88\linewidth]{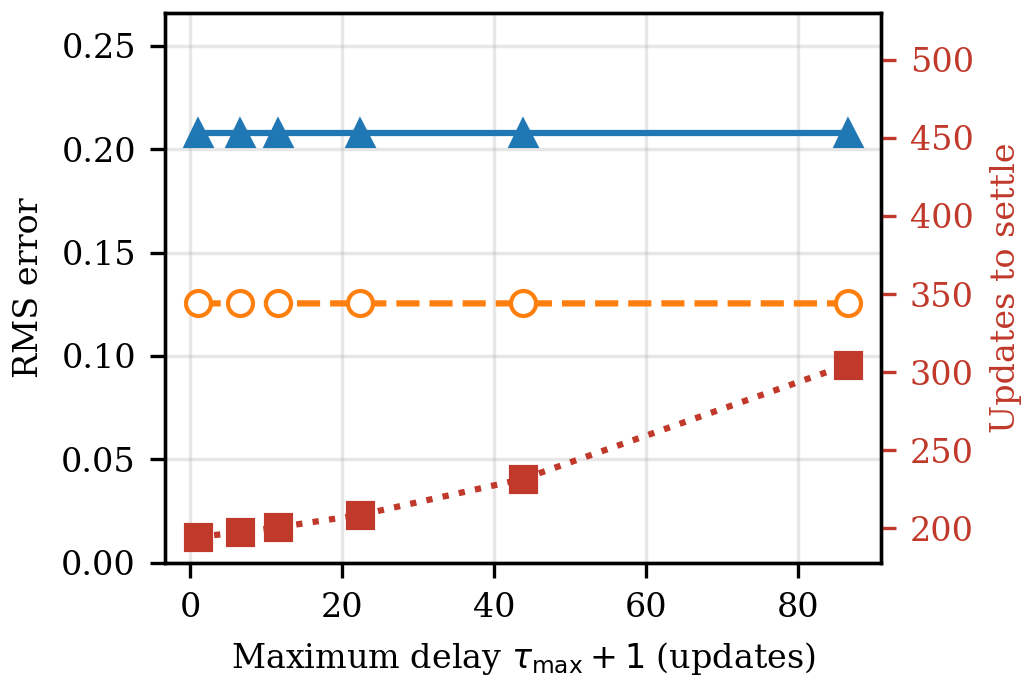}\label{fig:quant_delay}}
    \caption{Effects of quantization and delay on fixed JENC configurations (10 instances).}
    \label{fig:quant}
\end{figure}

\subsection{Effects of Quantization, Delay, and Sensing Uncertainty}\label{sec:quant}
Fig.~\ref{fig:quant_error} and~\ref{fig:quant_comm} use the same resolution $b$ in all links of fixed uncapped configurations.
The bound and the realized error decrease as $b$ increases. The bound holds in all $80$ evaluations.
At 2 bits, the bound is $0.493$, while the realized error is $0.124$. The communication bound exceeds the realized deviation by a median factor of $4.3$.
Therefore, the bound is more conservative at lower resolutions.

To verify Lemma~\ref{lem:wc} for delay, Fig.~\ref{fig:quant_delay} scales all link delays of fixed 45-kbps configurations by $s\in\{0,0.25,0.5,1,2,4\}$, so that the mean maximum delay reaches $85.6$ updates.
The bound remains $0.208$, because its steady-state terms do not depend on the delays. The realized error remains $0.125$ at all scales.
However, the lossless settling time increases from $195$ updates at zero delay to $305$ updates at $s=4$, as predicted by~\eqref{eq:transient}.

Fig.~\ref{fig:uncertainty_noise} scales all sensor standard deviations of the 45-kbps instances by a factor from 0.25 to 4, and \ac{jenc} redesigns the network for each scale.
The realized error grows almost linearly with the noise scale, from $0.033$ to $0.412$, and the bound holds in all evaluations.
However, the communication terms of the bound slightly change because they depend on the link resolutions and the gains but not on the measurements.
Therefore, the communication terms account for $67\%$ of the bound at the smallest noise scale, but only $17\%$ at the largest scale.
In other words, accurate sensors make the error communication-limited, and the network configuration matters most, whereas noisy sensors make it measurement-limited.

Fig.~\ref{fig:uncertainty_spread} varies the heterogeneity of the sensors while keeping the geometric mean of $\sigma_n$ fixed.
With identical sensors, inverse-variance and homogeneous gains coincide.
As $\sigma_{\max}/\sigma_{\min}$ increases to $21$, the realized error of homogeneous gains increases to $0.528$, whereas that of inverse-variance gains decreases to $0.096$, i.e., it is $79.3\%$ lower.
This is because inverse-variance gains rely on the most accurate sensors, as predicted by Proposition~\ref{prop:structure}(a).

\subsection{Components of JENC}

Table~\ref{tab:ablation} examines the components of \ac{jenc}.
Without the communication proxy $D_i$, the endpoint scores ignore the quantization load, and the bound increases from $0.142$ to $0.337$.
Adding the information proxy $F_i$ with $w_f=10$ slightly increases the bound and the realized error, consistent with the tuned value $w_f=5$.
Without coordinator verification, the final error is lower ($0.060$ versus $0.083$).
However, the bound increases during accepted actions in $80.0\%$ of runs, whereas it never increases in the other variants.
Moreover, $168.9$ accepted actions are required instead of $28.9$.
Therefore, the verification trades part of the realized error for fewer reconfigurations and the guarantees of Proposition~\ref{prop:auction}.
Without the rate-first pass, the bound and the realized error increase to $0.164$ and $0.094$, because an accepted shortcut reduces $\bar\alpha$ and cannot be undone by a single action.

\begin{table}[t]
    \centering
    \footnotesize
    \caption{Component analysis of JENC over 20 instances (45~kbps)}
    \label{tab:ablation}
    \resizebox{\columnwidth}{!}{%
    \begin{tabular}{lccc}
        \toprule
        Variant & $\etot$ & $\Etot$ & Accepted actions \\
        \midrule
        JENC & $0.142$ & $0.083$ & $28.9$ \\
        With information term ($w_f=5$) & $0.141$ & $0.084$ & $28.6$ \\
        Without communication proxy ($w_d=0$) & $0.337$ & $0.141$ & $43.3$ \\
        Without coordinator verification & $0.111$ & $0.060$ & $168.9$ \\
        Without rate-first pass & $0.164$ & $0.094$ & $15.3$ \\
        \bottomrule
    \end{tabular}}
\end{table}

Finally, at the output of \ac{jenc} in the main study, an exact evaluation of all feasible single-link actions finds no remaining improving action in $117$ of $120$ configurations, and the largest remaining decrease of $J$ is $5.0\%$.
Additionally, on $8$ six-agent instances with exhaustive search, the mean and maximum optimality gaps of \ac{jenc} are $1.4\%$ and $3.7\%$, compared with $22.9\%$ and $97.7\%$ for GNC, respectively.

\section{Conclusion}\label{sec:conclusion}
This paper presented a joint estimation and network configuration framework for multi-agent sensing under delayed and finite-rate communication.
A computable bound on the estimation error that separates the disagreement from the inaccuracy is derived.
Then, a semi-distributed algorithm that uses this bound to configure the communication graph, the link rates, and the gains while preserving feasibility and descent is presented.
The evaluation showed that the bound holds and the estimation error is reduced in all tested networks.
Consequently, the realized estimation error is reduced with fewer links and topology changes compared with GNC, L2C, and RLC. In addition, less computation is needed compared with other bound-driven methods.
However, the estimation bound is conservative and ranks configurations weakly under tight budgets.
Future work will tighten the communication bound by accounting for the averaging of quantization errors, add multi-link actions, and extend the framework to fresh measurements, time-varying states, and mobile agents.

\appendix

\subsection*{Proof of Lemma~\ref{lem:fixed}}
\label{proof:1}
The matrix $B+\alpha L$ has positive diagonal entries $\beta_n+\alpha d_n$ and nonpositive off-diagonal entries. It is strictly row diagonally dominant because $\beta_n>0$. Hence, it is a nonsingular M-matrix, and $G=(B+\alpha L)^{-1}\ge0$~\cite{berman1994nonnegative}, which implies $H=GB\ge0$. Since $L\one=\bm 0$, $(B+\alpha L)\one=B\one$, thus $H\one=GB\one=\one$. Since $\one^\top L=\bm 0^\top$, $\one^\top(B+\alpha L)=\betab^\top$, so that $\one^\top=\betab^\top G$ and $\betab^\top H=\betab^\top GB=\one^\top B=\betab^\top$.

An equilibrium point of the ideal recursion satisfies $\bm P^\circ=(W\otimes I_r)\bm P^\circ+(B\otimes I_r)\tilde{\bm P}$, i.e., $((B+\alpha L)\otimes I_r)\bm P^\circ=(B\otimes I_r)\tilde{\bm P}$, whose unique solution is $(H\otimes I_r)\tilde{\bm P}$.

Since the rows of $H$ are nonnegative and its sum is one, each $\pfix_n$ is a convex combination of the measurements. Premultiplying by $\betab^\top\otimes I_r$ and using $\betab^\top H=\betab^\top$ yields~\eqref{eq:weighted_mean}. Under~\eqref{eq:gains}, $\beta_n\propto\sigma_n^{-2}$. Therefore, $\pbeta=\bm p_{\mathrm{BLUE}}$. The convergence under bounded delays follows from~\eqref{eq:transient} with $\bm q=\bm 0$. \hfill$\blacksquare$

\subsection*{Proof of Lemma~\ref{lem:wc}}
\label{proof:2}
Let $\bm e_{nm}[k]=\bm q_{nm}[k]-\phat_m[k-\tau_{nm}]$. Subtracting the equilibrium point $\pfix_n=(1-\beta_n-\alpha d_n)\pfix_n+\alpha\sum_ma_{nm}\pfix_m+\beta_n\ptil_n$ from~\eqref{eq:estimator} gives the exact error recursion
\begin{equation}
\begin{aligned}
    \bm x_n[k+1]={}&(1-\beta_n-\alpha d_n)\bm x_n[k]\\
    &+\alpha\textstyle\sum_ma_{nm}\big(\bm x_m[k-\tau_{nm}]+\bm e_{nm}[k]\big).
\end{aligned}
    \label{eq:error_recursion}
\end{equation}
With Assumption~\ref{ass:step}, all coefficients are nonnegative and sum to $1-\beta_n$. With $\Psi[k]=\max_{n,\,k-\tau_{\max}\le j\le k}\|\bm x_n[j]\|$ and~\eqref{eq:quantization}, the triangle inequality yields $\|\bm x_n[k+1]\|\le(1-\bmin)\Psi[k]+q_{\max}$, where $q_{\max}=\max_nq_n$. Let $\bar r=q_{\max}/\bmin$ and $M_i=\bar r+(1-\bmin)^i\max(\psi_0-\bar r,0)$, so that $M_{i+1}=(1-\bmin)M_i+q_{\max}\le M_i$. We claim that $\|\bm x_n[j]\|\le M_i$ for all $n$ and all $j\ge k_i$, where $k_0=-\tau_{\max}$ and $k_i=1+(i-1)(\tau_{\max}+1)$ for $i\ge1$. For $i=0$, $\Psi[0]=\psi_0\le M_0$, and whenever every value in a window is at most $M_0$, the subsequent value is at most $M_1\le M_0$. So, no value exceeds $M_0$. If the claim holds for $i$, then for every $k\ge k_i+\tau_{\max}$ the window $[k-\tau_{\max},k]$ lies in $[k_i,\infty)$, and therefore $\|\bm x_n[k+1]\|\le M_{i+1}$ for all $k+1\ge k_i+\tau_{\max}+1=k_{i+1}$ (for $i=0$, this yields $k_1=1$). For $k\ge1$ and $i=\lfloor k/(\tau_{\max}+1)\rfloor$, we have $k\ge k_i$ and $M_i\le\bar r+(1-\bmin)^i\psi_0$, which establishes~\eqref{eq:transient}.

Consequently, $\xi_n=\limsup_k\|\bm x_n[k]\|$ is finite. Taking the $\limsup$ of the norm of~\eqref{eq:error_recursion} and using $\limsup_k\|\bm x_m[k-\tau_{nm}]\|=\xi_m$ yields $\xi_n\le(1-\beta_n-\alpha d_n)\xi_n+\alpha\sum_ma_{nm}\xi_m+q_n$, i.e., $(B+\alpha L)\bm\xi\le\bm q$. Premultiplying by $G\ge0$ results in~\eqref{eq:xbar_wc}. \hfill$\blacksquare$

\subsection*{Proof of Theorem~\ref{thm:total}}
\label{proof:3}
Rewrite $\phat_n=\pfix_n+\bm x_n$, $\bar{\bm p}^\circ=\tfrac1N\sum_m\pfix_m$, and $\bar{\bm x}=\tfrac1N\sum_m\bm x_m$, and let $\nu_n=\limsup_k(\E\|\bm x_n[k]\|^2)^{1/2}$. By~\eqref{eq:transient}, $\|\bm x_n[k]\|^2\le(\max_mq_m/\bmin+\psi_0)^2$ for $k\ge1$, where $\psi_0=\max_n\|\bm v_n-\sum_mH_{nm}\bm v_m\|\le2(\sum_m\|\bm v_m\|^2)^{1/2}$ is square integrable by~\eqref{eq:obs_model}. Therefore, the reverse Fatou lemma~\cite{stein2009real} and~\eqref{eq:xbar_wc} yield $\limsup_k\E\|\bm x_n[k]\|^2\le\E\limsup_k\|\bm x_n[k]\|^2\le\xbar_n^2$. Hence, $\nu_n\le\xbar_n$.

\emph{Disagreement:} Since $\phat_n-\pbar=(\pfix_n-\bar{\bm p}^\circ)+(1-\tfrac1N)\bm x_n-\tfrac1N\sum_{m\ne n}\bm x_m$, the Minkowski inequality at each $k$, together with the subadditivity of $\limsup$, yields $\limsup_k(\E\|\phat_n-\pbar\|^2)^{1/2}\le(\E\|\pfix_n-\bar{\bm p}^\circ\|^2)^{1/2}+(1-\tfrac2N)\xbar_n+\tfrac1N\sum_m\xbar_m$. Since $\bm P^\circ-\one\otimes\pstar=(H\otimes I_r)\bm V$, where $\bm V$ stacks the noise vectors, $\pfix_n-\bar{\bm p}^\circ=\sum_m[\Pi H]_{nm}\bm v_m$, and independence together with~\eqref{eq:obs_model} results in $\E\|\pfix_n-\bar{\bm p}^\circ\|^2=r[\Pi HSH^\top\Pi]_{nn}=\epsilon_{\mathrm{dis},n}^2$. Maximizing over $n$ yields $\Eagr\le\tfrac12\eest$, and~\eqref{eq:pairwise} yields $\Eest\le\eest$.

\emph{Inaccuracy:} Decompose $\pbar-\pstar=\bar{\bm x}+(\pbeta-\pstar)+(\bar{\bm p}^\circ-\pbeta)$, where $\pbeta-\pstar=\sum_m\frac{\beta_m}{\sum_j\beta_j}\bm v_m$ and $\bar{\bm p}^\circ-\pbeta=\sum_mw_m\bm v_m$. Their second moments are equal to $\epsilon_{\mathrm{fus}}^2$ and $\epsilon_{\mathrm{gs}}^2$, respectively, and $\limsup_k(\E\|\bar{\bm x}\|^2)^{1/2}\le\tfrac1N\sum_m\xbar_m$. The Minkowski inequality yields $\Eacc\le\eacc$, and~\eqref{eq:split} yields $\Etot\le\etot$. \hfill$\blacksquare$

\subsection*{Proof of Proposition~\ref{prop:structure}}
\label{proof:4}
(a) By Cauchy--Schwarz Inequality, $(\sum_n\beta_n)^2=(\sum_n\beta_n\sigma_n\cdot\sigma_n^{-1})^2\le\sum_n\beta_n^2\sigma_n^2\sum_n\sigma_n^{-2}$, with equality if and only if $\beta_n\sigma_n\propto\sigma_n^{-1}$. Rearranging results in $\epsilon_{\mathrm{fus}}^2\ge r/\sum_n\sigma_n^{-2}=\epsilon_{\mathrm{BLUE}}^2$.

(b) By~\eqref{eq:weighted_mean}, $\sum_n\beta_n\pfix_n=N\bar\beta\pbeta$. Moreover, $\sum_n\beta_n\pfix_n=N\bar\beta\bar{\bm p}^\circ+\sum_n(\beta_n-\bar\beta)(\pfix_n-\bar{\bm p}^\circ)$, because $\sum_n(\beta_n-\bar\beta)=0$.
Combining the two expressions and rearranging provides the network average difference for the given gain deviation.
The Cauchy--Schwarz inequality gives $\|\bar{\bm p}^\circ-\pbeta\|\le\frac{\|\betab-\bar\beta\one\|}{N\bar\beta}\big(\sum_n\|\pfix_n-\bar{\bm p}^\circ\|^2\big)^{1/2}$, and taking second moments yields~\eqref{eq:gain_spread}. For homogeneous gains, $H$ is symmetric. Therefore, $H^\top\one=\one$ and $\bm w=\bm 0$.

(c) From the proof of Lemma~\ref{lem:fixed}, $\betab^\top G=\one^\top$, so that $\betab^\top\xbar=\one^\top\bm q$. Since $\xbar\ge0$, $\bmax\sum_n\xbar_n\ge\sum_n\beta_n\xbar_n=N\bar\epsilon_c$. \hfill$\blacksquare$

\subsection*{Proof of Proposition~\ref{prop:scale}}
\label{proof:5}
Let $H_{\beta_0}=(\beta_0\tilde W+\alpha L)^{-1}\beta_0\tilde W$ with $\tilde W=\mathrm{diag}(\tilde{\bm w})$. Each $H_{\beta_0}$ is nonnegative and row stochastic. Therefore, it is bounded and satisfies $\alpha LH_{\beta_0}=\beta_0\tilde W(I-H_{\beta_0})\to0$. Thus, every limit point $H_0$ satisfies $LH_0=0$. By connectivity, its columns lie in $\mathrm{span}(\one)$, so that $H_0=\one\bm h^\top$ with $\bm h^\top\one=1$. Since $\tilde{\bm w}^\top H_{\beta_0}=\tilde{\bm w}^\top$ by Lemma~\ref{lem:fixed}, $\bm h=\tilde{\bm w}/(\one^\top\tilde{\bm w})$. The limit point is unique, so $H_{\beta_0}$ converges to it.

Then, $\Pi H_{\beta_0}\to0$ and $\bm w\to\bm 0$, so that $\epsilon_{\mathrm{dis},n}\to0$ and $\epsilon_{\mathrm{gs}}\to0$. Moreover, $\epsilon_{\mathrm{fus}}$ is invariant with respect to scaling of $\betab$. This proves (a).

For (b), Proposition~\ref{prop:structure}(c) applies with $\bmax=\beta_0$, and $\bar\epsilon_c>0$ does not depend on $\beta_0$. Hence, $\etot\ge\eacc\ge\bar\epsilon_c/\beta_0$.

The bound $\etot$ is continuous on $(0,\beta_0']$ and diverges as $\beta_0\to0$. Its minimum is therefore attained at a positive scale in that admissible range. \hfill$\blacksquare$

\subsection*{Proof of Corollary~\ref{cor:spectral}}
\label{proof:6}
Let $\bm\delta^\circ=(\Pi\otimes I_r)\bm P^\circ$. Applying $\Pi$ to the equilibrium-point equation and using $\Pi W\one=-(\betab-\bar\beta\one)$ and $\sum_n\beta_n(\ptil_n-\pbeta)=\bm 0$ yields $\bm\delta^\circ=(\Pi W\Pi\otimes I_r)\bm\delta^\circ+\bm\varsigma+(\betab-\bar\beta\one)\otimes(\pbeta-\bar{\bm p}^\circ)$, where $\bm\varsigma=(B\otimes I_r)(\tilde{\bm P}-\one\otimes\pbeta)$. Since $\bm\delta^\circ\in\one^\perp\otimes\mathbb R^r$, $\|(\Pi W\Pi\otimes I_r)\bm\delta^\circ\|\le\rho\|\bm\delta^\circ\|$, and Proposition~\ref{prop:structure}(b) gives $\|\pbeta-\bar{\bm p}^\circ\|\le\frac{g}{N\bar\beta}\|\bm\delta^\circ\|$.

Hence, $\Delta\|\bm\delta^\circ\|\le\bar\beta\|\bm\varsigma\|$ for every realization, and taking second moments yields~\eqref{eq:spectral}, because $\sum_n\epsilon_{\mathrm{dis},n}^2=\E\|\bm\delta^\circ\|^2$ and $\E\|\bm\varsigma\|^2=\bar d_s^2$. For homogeneous gains, the eigenvalues of $U^\top WU$ are $1-\beta-\alpha\lambda_i(L)$, $i\ge2$. \hfill$\blacksquare$

\subsection*{Proof of Proposition~\ref{prop:auction}}
\label{proof:7}
(a) The initial state is verified to be feasible, and every accepted action is re-checked against the budgets and $\lambda_2\ge\underline\lambda_2>0$, which implies connectivity. The deployed gain belongs to $\mathcal A(L)\subset(0,\bar\alpha]$.

(b) Every accepted action decreases $J$ by at least $\delta$, and $J\ge\etot\ge\epsilon_{\mathrm{fus}}$, which is independent of the configuration because $\betab$ is fixed. A strictly decreasing potential cannot revisit a configuration.

(c) A sweep without accepted actions leaves the state unchanged. Every feasible score-qualified proposal was therefore evaluated at that state and violated at least one acceptance test. When $\mu=0$, the nonincreasing-bound test is redundant, so each such proposal violated the $\delta$-descent test.

\bibliographystyle{IEEEtran}
\bibliography{biblo}

\end{document}